\documentstyle[12pt]{article}
\parskip=\medskipamount                 % space between paragraphs
\thicklines                     % thick straight lines for pictures
\newcommand{\bim}[6]{\bibitem{#1}#2, {\em #3\/}$\;${\bf
#4}$\;$(#5)$\;${#6}.}
\def\IR{\relax{\rm I\kern-.18em R}}
\def\ZZ{\relax{\sf Z\kern-.4em Z}}
\def\a{\alpha} \def\b{\beta}    
 \def\l{\lambda} 
   
  \def\cC{{\cal C}} \def\cD{{\cal D}}
   
 \def\cK{{\cal K}} \def\cL{{\cal L}} \def\cM{{\cal M}}
 \def\cO{{\cal O}}  
 
\newtheorem{proposition}{Proposition}[section]
\newtheorem{corollary}{Corollary}[section]
\newtheorem{conjecture}{Conjecture}[section]

\newlength{\shiftwidth}
\def\shift#1{&&\hbox to \shiftwidth{\hfill $\displaystyle#1$}}
\newlength{\sshiftwidth}
\def\sshift#1{\lefteqn{\hbox to
\sshiftwidth{\hfill$\displaystyle#1$}}}
\def\lshift#1{\hbox to \shiftwidth{$\displaystyle#1$}\hfill}

\def\sign#1{{\rm sign}\left(#1\right)}

\def\ord{{\rm ord}\,}

\def\Tr{{\rm Tr}}
\def\Pexp{{\rm Pexp}}
\def\hol#1{\Pexp\left(\oint_{#1} A_\mu dx^\mu\right)}
\def\thol#1#2{\Tr_{#1}\hol{#2}}

\def\Tub{{\rm Tub}}

\def\ztramk{Z^{({\rm tr})}_{\a}(M,\cK;k)}

\def\ztrmk{Z^{({\rm tr})}(M;k)}

\def\rhs{RHS$\;$}

\def\ohm{\ord H_1(M,\ZZ)}

\catcode`\@=11

\newif\if@fewtab\@fewtabtrue

\catcode`\@=11

\newif\if@fewtab\@fewtabtrue

{\count255=\time\divide\count255 by 60
\xdef\hourmin{\number\count255}
\multiply\count255 by-60\advance\count255 by\time
\xdef\hourmin{\hourmin:\ifnum\count255<10 0\fi\the\count255}}
\def\ps@draft{\let\@mkboth\@gobbletwo
    \def\@oddhead{}
    \def\@oddfoot
       {\hbox to 7 cm{$\scriptstyle Draft\ version:\ \draftdate$
       \hfil}\hskip -7cm\hfil\rm\thepage \hfil}
    \def\@evenhead{}\let\@evenfoot\@oddfoot}

\def\ceqno{\global\@fewtabfalse
    \ifcase\@eqcnt \def\@tempa{& & &}\or \def\@tempa{& &}
      \or \def\@tempa{&}
      \or\def\@tempa{}\fi\@tempa
{\rm(\theequation)}}

\def\aeqno#1{\global\@fewtabfalse
    \ifcase\@eqcnt \def\@tempa{& & &}\or \def\@tempa{& &}
      \or \def\@tempa{&}
      \or\def\@tempa{}\fi\@tempa
{\rm(\theequation,#1)}}

\def\label#1{\ifnum\draftcontrol=1
 \global\def\draftnote{$\scriptstyle #1$}\fi
 \@bsphack\if@filesw {\let\thepage\relax
   \def\protect{\noexpand\noexpand\noexpand}%
\xdef\@gtempa{\write\@auxout{\string
      \newlabel{#1}{{\@currentlabel}{\thepage}}}}}\@gtempa
   \if@nobreak \ifvmode\nobreak\fi\fi\fi
  \@esphack}

\def\alabel#1#2{\label{#1}\global\@fewtabfalse
    \ifcase\@eqcnt \def\@tempa{& & &}\or \def\@tempa{& &}
      \or \def\@tempa{&}
      \or\def\@tempa{}\fi\@tempa
{\hbox to 3cm{\phantom{\rm(\theequation,#2)}
\draftnote \hfil}\hskip -3cm {\rm(\theequation,#2)}}}

\def\clabel#1{\label{#1}\global\@fewtabfalse
    \ifcase\@eqcnt \def\@tempa{& & &}\or \def\@tempa{& &}
      \or \def\@tempa{&}
      \or\def\@tempa{}\fi\@tempa
{\hbox to 3cm{\phantom{\rm(\theequation)}
\draftnote \hfil}\hskip -3cm{\rm(\theequation)}}}

\def\eqnarray{\def\draftnote{{}}\global\@fewtabtrue
\stepcounter{equation}\let\@currentlabel=\theequation
\global\@eqnswtrue
\global\@eqcnt\z@\tabskip\@centering\let\\=\@eqncr
$$\halign to \displaywidth\bgroup\@eqnsel\hskip\@centering\@eqcnt\z@
  $\displaystyle\tabskip\z@{##}$&\global\@eqcnt\@ne
  \hskip 1\arraycolsep \hfil${##}$\hfil
  &\global\@eqcnt\tw@ \hskip 1\arraycolsep
$\displaystyle\tabskip\z@{##}$
\hfil  \tabskip\@centering&\global\@eqcnt\thr@@\llap{##}\tabskip\z@
\cr}

\def\endeqnarray{\@@eqncr\egroup
      \global\advance\c@equation\m@ne$$\global\@ignoretrue}

\def\@eqnnum{\hbox to 3cm{\phantom{\rm(\theequation)} \draftnote
                         \hfil}\hskip -3cm {\rm(\theequation)}}

\def\@@eqncr{\let\@tempa\relax
    \ifcase\@eqcnt \def\@tempa{& & &}\or \def\@tempa{& &}
      \or \def\@tempa{&}
      \or\def\@tempa{}
\fi\@tempa
\if@eqnsw
\if@fewtab\@eqnnum\fi
\stepcounter{equation}\fi\global
\@eqnswtrue\global\@eqcnt\z@\global\@fewtabtrue\cr}

\def\draftcite#1{\ifnum\draftcontrol=1#1\else{}\fi}

\def\@lbibitem[#1]#2{\item{}\hskip -3cm \hbox to 2cm
{\hfil$\scriptstyle\draftcite{#2}$}\hskip
1cm[\@biblabel{#1}]\if@filesw
     {\def\protect##1{\string ##1\space}\immediate
      \write\@auxout{\string\bibcite{#2}{#1}}}\fi\ignorespaces}

\def\@bibitem#1{\item\hskip -3cm \hbox to 2cm
{\hfil $\scriptstyle\draftcite{#1}$}\hskip 1cm
\if@filesw \immediate\write\@auxout
       {\string\bibcite{#1}{\the\value{\@listctr}}}\fi\ignorespaces}

\def\nsection#1{\section{#1}\setcounter{equation}{0}}
\def\draftdate{\number\month/\number\day/\number\year\ \ \ \hourmin }

\global\def\draftcontrol{0}
\catcode`\@=12

\def\theequation{{\thesection.\arabic{equation}}}

\def\qq{\begin{eqnarray}}
\def\qqq{\end{eqnarray}}

\begin{document}
%\draft

\begin{titlepage}
\centerline{\hfill                 UMTG-175-94}
\centerline{\hfill                 hep-th/9403020}
\vfill
\begin{center}
{\large \bf
Reshetikhin's Formula for the Jones Polynomial of a Link: Feynman
Diagrams and Milnor's Linking Numbers.
} \\

\bigskip
\centerline{L. Rozansky\footnote{Work supported
by the National Science Foundation
under Grant No. PHY-92 09978.
}}

\centerline{\em Physics Department, University of Miami
}
\centerline{\em P. O. Box 248046, Coral Gables, FL 33124, U.S.A.}

\vfill
{\bf Abstract}

\end{center}
\begin{quotation}

   We use Feynman diagrams to prove a formula for the Jones polynomial
of a link derived recently by N.~Reshetikhin. This formula presents
the colored Jones polynomial as an integral over the coadjoint orbits
corresponding to the representations assigned to the link components.
The large $k$ limit of the integral can be calculated with the help
of the stationary phase approximation. The Feynman rules allow us to
express the phase in terms of integrals over the manifold and the
link components. Its stationary points correspond to flat connections
in the link complement. We conjecture a relation between the dominant
part of the phase and Milnor's linking numbers. We check it explicitly
for the triple and quartic numbers by comparing their expression
through the Massey product with Feynman diagram integrals.

\end{quotation}
\vfill
\end{titlepage}

\pagebreak
%\tableofcontents
%\pagebreak
%%%%%%%%%%%%%% MORE DEFINITIONS %%%%%%%%%%%%%%
\def\vga{\vec{\a}}
\def\va{\vec{a}}
\def\vb{\vec{b}}
\def\vgb{\vec{\b}}
\def\vgr{\vec{\rho}}
\def\vx{\vec{x}}
\def\vn{\vec{n}}
\def\vv{\vec{v}}
\def\vA{\vec{A}}
\def\vC{\vec{C}}
\def\vgs{\vec{\sigma}}

\def\ztraml{Z^{({\rm tr})}_{\a_1,\ldots,\a_n}(M,\cL,k)}
\def\zasl{Z_{\a_1,\ldots,a_n}(S^3,\cL;k)}
\def\ztrasl{Z^{({\rm tr})}_{\a_1,\ldots,\a_n}(S^3,\cL;k)}

\def\pva{\prod_{j=1}^{n}\left(\frac{K}{4\pi}\frac{d^2\va_j}{|\va_j|}\right)}
\def\spint{\int_{|\va_j|=\frac{\a_j}{K}}\pva}
\def\adots{(\va_1,\ldots,\va_n)}
\def\lm{L_m\adots}
\def\elm{\exp\left(\frac{i\pi K}{2}\sum_{m=2}^{\infty}\lm\right)}
\def\pml{P_{m,l}\adots}
\def\spml{\sum_{\stackrel{\scriptstyle l,m=0}{l+m\neq
0}}^{\infty}K^{-m}\pml}
\def\onespml{\left[1+\spml\right]}
\def\finv{F_m(\vb_1,\ldots,\vb_m)}
\def\lt{l^{(3)}}
\def\lti{\lt_{ijk}}
\def\lf{l^{(4)}}
\def\lfi{\lf_{ij,kl}}
\def\lmi{l^{(\mu)}_{i_1,\ldots,i_m}}
\def\lmij{l^{(\mu)}_{i_1,\ldots,i_{m-1},j}}

\def\eadots{e^{2\pi ia_1},\ldots,e^{2\pi ia_n}}
\def\mintub{M\setminus{\rm Tub}(\cL)}
\def\dal{\Delta_A(M,\cL;\eadots)}
\def\invrtl{\tau_R^{-1}(\mintub;\eadots)}
\def\mind{M_{ij,\mu\nu}}
\def\emind{e^{\frac{i\pi}{4}\sign{\mind}}}
\def\emmind{e^{-\frac{i\pi}{4}\sign{\mind}}}
\def\pl{P_{0,l}(a_1\vn,\ldots,a_n\vn)}
\def\pln{P_{0,l}(a_1\vn,\ldots,a_{n-1}\vn,0)}

\def\spl{\sum_{l=2}^{\infty}\pl}
\def\spln{\sum_{l=2}^{\infty}\pln}
\def\proda{\prod_{j=1}^{n}a_j}
\def\addots{a_1,\ldots,a_n}

\def\fpq{\frac{p}{q}}
\def\pqlf{\fpq+l_{nn}}
\def\pql{p+ql_{nn}}
\def\fpql{\frac{q}{\pql}}
\def\emsp{e^{-\frac{3}{4}i\pi\sign{\pqlf}}}
\def\skq{\sqrt{2K|q|}}
\def\fsq{\frac{2\sign{q}}{\skq}}

\def\cv{C(v_1,\ldots,v_n)}
\def\cvn{C_n(v_1,\ldots,v_n)}
\def\cvna{C_n^a(v_1,\ldots,v_n)}
\def\caa{C^a_{a_1,\ldots,\a_n}}
\def\p{^\prime}
\def\pp{^{\prime\prime}}

%%%%%%%%% END OF MORE DEFINITIONS %%%%%%%%%%%%%%%%%%%%%%%

%+++++++++++++++++++++++++++++++++++++++++++++++++++
\nsection{Introduction}
%+++++++++++++++++++++++++++++++++++++++++++++++++++

Let $\cL$ be an $n$-component link in a 3-dimensional manifold $M$.
E.~Witten presented in~\cite{Wi1}
the Jones polynomial of $\cL$ as a path
integral over the gauge equivalence classes of $SU(2)$ connection
$A_\mu$ on $M$:
\qq
Z_{\a_1,\ldots,\a_n}(M,\cL;k)=\int[\cD A_\mu]
\exp\left(\frac{i}{\hbar}S_{CS}\right)
\prod_{j=1}^{n}\thol{\a_j}{\cL_j},
\label{1.1}
\qqq
here $S_{CS}$ is the Chern-Simons action
\qq
S_{CS}=\frac{1}{2}\,\Tr\,\epsilon^{\mu\nu\rho}\int_{M}d^3x\;
(A_\mu \partial_\nu A_\rho - \frac{2}{3}A_\mu A_\nu A_\rho),
\label{1.2}
\qqq
$\hbar$ is a ``Planck's constant'':
\qq
\hbar=\frac{2\pi}{k},\;\;k\in \ZZ,
\label{1.3}
\qqq
the trace $\Tr$ in eq.~(\ref{1.2})
is taken in the fundamental (2-dimensional)
representation and $\thol{\a_j}{\cL_j}$ are the traces of holonomies
along the link components $\cL_j$ taken in the
$\a_j$-dimensional representations.

The path integral~(\ref{1.1}) can be calculated in the stationary
phase approximation in the limit of large $k$. The stationary points
of the Chern-Simons action~(\ref{1.2}) are flat connections and
Witten's invariant is presented as a sum over connected pieces $\cM_c$
of their moduli space $\cM$:
\begin{eqnarray}
Z_{\a_1,\ldots,\a_n}(M,\cL;k)&=&
\sum_{\cM_c}Z^{(\cM_c)}_{\a_1,\ldots,\a_n}(M,\cL;k),
\nonumber\\
Z^{(\cM_c)}_{\a_1,\ldots,\a_n}(M,\cL;k)&=&
\exp\frac{i}{\hbar}\left(S_{CS}^{(c)}+\sum_{n=1}^{\infty}
S_n^{(c)}\hbar^n\right),
\label{1.13}
\end{eqnarray}
here $S_{CS}$ is a Chern-Simons action of flat connections of $\cM_c$
and $S_n^{(c)}$ are the quantum $n$-loop
corrections to the
contribution of $\cM_c$.

Suppose that $M$ is a rational homology sphere (\rhs). Then the
trivial connection is an isolated point in the moduli space of flat
connections. Let $\cL$ have only one component, so that it is a knot
$\cK$. In our previous paper~\cite{RoI} we gave a ``path integral''
proof of the following conjecture which P.~Melvin and H.~Morton
formulated in~\cite{MeMo} for the case of $M=S^3$:
%%%%%%%%%%%%%%%%%%%%%%%%%%%%%%%%%%
\begin{proposition}
\label{pmy.1}
The trivial connection contribution to the Jones polynomial of a knot
$\cK$ in a \rhs $M$ can be expressed as
\qq
\ztramk =\ztrmk\,\exp\left[\frac{i\pi}{2K}\nu(\a^2-1)\right]
\,\a J(\a,K),
\label{2.15}
\qqq
here $\nu$ is a self-linking number of $\cK$ and $J(\a,K)$ is a
function that has the following expansion in $K^{-1}$ series:
\qq
J(\a,K)=\sum_{n=0}^{\infty}\sum_{m=0}^{n}
D_{m,n}\a^m K^{-n}.
\label{2.16}
\qqq
The dominant part of this expansion is related to the Alexander
polynomial of $\cK$:
\qq
\pi a\sum_{n=0}^{\infty}D_{n,n}a^n=
[\ohm]\frac{\sin\left(\frac{\pi a}{m_2 d}\right)}
{\Delta_A \left(M,\cK;e^{2\pi i\frac{a}{m_2 d}}\right)},
\label{my.2}
\qqq
the integer numbers $m_2$ and $d$ are defined in~\cite{RoI}, $m_2=d=1$
if $M=S^3$.
\end{proposition}
%%%%%%%%%%%%%%%%%%%%%%%%%%%%%%%%%%%%%%%%%%%%%%%
%

The virtue of eq.~(\ref{2.15}) is that it assembles the dominant part
of the $1/K$ expansion of the Jones polynomial $\ztramk$ into the
exponential and puts a restriction on the power of $\a$ in the
preexponential factor. This enabled us in~\cite{RoI} to use a
stationary phase approximation in the Witten-Reshetikhin-Turaev
surgery formula in order to derive a knot surgery formula for the
loop corrections $S_n^{({\rm tr})}$. A similar formula for the Jones
polynomial of a link is required in order to generalize this result
to link surgery. However the arguments of~\cite{RoI} which led to
eq.~(\ref{2.15}) can not be extended directly to links, because there
may exist irreducible flat connections in the link complement with
arbitrarily small holonomies along the meridians of link components.

A generalization of eq.~(\ref{2.15}) for links was
derived recently by N.~Reshetikhin\footnote{I am indebted to
N.~Reshetikhin for sharing the results of his unpublished
research.}~\cite{Re}. He observed that if the dimensions $\a_i$ in
eq.~(\ref{1.1}) are big enough, then the representation spaces can be
treated classically: the matrix elements of Lie algebra generators in
$\a_j$-dimensional representation can be substituted by functions on
the coadjoint orbit of radius $\a_j$ and a trace over the
representation can be substituted by an integral over that orbit.
%%%%%%%%%%%%%%%%%%%%%%%%%%%%%%%%%%%%%
\begin{proposition}
\label{pf2.1}
Let $\cL$ be an $n$-component link in a \rhs $M$. Then the trivial
connection contribution to its Jones polynomial can be expressed as a
multiple integral over the $SU(2)$ coadjoint orbits:
\begin{eqnarray}
\lefteqn{\ztraml=\ztrmk\spint\elm}
\label{5.1}\\
\shift{
\times\onespml.}
\nonumber
\end{eqnarray}
here $\va_j$ are 3-dimensional vectors with fixed length
\qq
|\va_j|=\frac{\a_j}{K}
\label{5.2}
\qqq
and $\lm$, $P_{l,m}\adots$ are homogeneous invariant (under $SO(3)$
rotations) polynomials of degree $M$. In particular,
\qq
L_2\adots=\sum_{i,j=1}^{n}l_{ij}\,\va_i\cdot\va_j,
\label{5.4}
\qqq
$l_{ij}$ is the linking number of the link components $\cL_i$ and
$\cL_j$.
\end{proposition}
%%%%%%%%%%%%%%%%%%%%%%%%%%%%%%%%%%%%%%%%%%%%%%%%%%%%%%%
An example of this formula for a torus link is derived in Appendix~1
of~\cite{RoII}.

In Section~\ref{*2} we find a set of Feynman rules to calculate the
trivial connection contribution to the Jones polynomial $\ztraml$.
This enables us to prove the Proposition~\ref{pf2.1} and to derive a
property (Proposition~\ref{pf.3}) of the polynomials $\lm$ that we
will use in~\cite{RoII} in order to relate the r.h.s. of
eq.~(\ref{5.1}) to the multivariable Alexander polynomial.
In Section~\ref{*3} we formulate a conjecture that the coefficients
of the polynomials $\lm$ are related to Milnor's linking
numbers\footnote{I am thankful to A.~Vaintrob and O.~Viro for
teaching me about these linking numbers.} in a way that generalizes
eq.~(\ref{5.4}), and present some circumstantial evidence that
supports it. In Section~\ref{*n1} we study the properties of flat
connections in the link complement in order to further support our
conjecture.
In Sections~\ref{*4}
and~\ref{*5} we use Feynman diagrams to calculate the coefficients of
the polynomials $L_m$ for $m=3,4$. We demonstrate explicitly that
they are indeed proportional to Milnor's triple and quartic linking
numbers.

We will use the following notations throughout this paper: an element
$v$ of the Lie algebra $su(2)$ can be presented as $v=i\sigma_a v^a$,
here $\sigma_a$ are $2\times 2$ Pauli matrices which form an
orthogonal basis in the fundamental representation of $SU(2)$:
$[\sigma_a,\sigma_b]=2i\epsilon_{abc}\sigma_c$,
$\Tr\sigma_a\sigma_b=2\delta_{ab}$. Three components $v^a$, $1\leq
a\leq 3$ form a 3-dimensional vector $\vv$. Depending on our needs,
we will use $v$, $v^a$ or $\vv$ to denote the same object.

%************************************************
\nsection{Feynman Diagrams}
\label{*2}
%************************************************

We are going to prove Reshetikhin's formula~(\ref{5.1}) for the Jones
polynomial of a link by using Feynman diagrams in order to calculate
the path integral~(\ref{1.1}). An introduction into Feynman diagram
technics can be found in any textbook on quantum field theory.
Feynman rules of the Chern-Simons theory~(\ref{1.2}) are described,
e.g. in~\cite{AxSi}. We also especially recommend~\cite{BN} for a
nice simple introduction into this subject.

Our goal is to ``take a logarithm'' of the sum of all Feynman
diagrams contributing to $\ztraml$. This would be easy if the link
$\cL$ had no components: the logarithm would be equal to the sum of
all connected Feynman diagrams. Therefore we have to find the analog
of ``connectivity'' for the diagrams with the end-points on the link
components. This will be the connectivity in quantum theory on
the coadjoint orbit which describes the $\a_j$-dimensional
representation of $SU(2)$ assigned to a link component $\cL_j$. The
key to defining the connectivity is the Campbell-Hausdorf
formula\footnote{I am indebted to N.~Reshetikhin for pointing to the
relevance of the Campbell-Hausdorf formula for his derivation of
eq.~(\ref{5.1}).} which shows how to take a logarithm of the product
of two noncommuting exponentials.

Let $\cK$ be a knot in a 3-dimensional manifold $M$. Let $A(t)$ be a
restriction of an $SU(2)$ connection $A_\mu$ onto $\cK$:
\qq
A(t)=A_\mu(x(t))\frac{dx^\mu}{dt},
\label{f.4}
\qqq
here $0\leq t\leq 1$ is a parametrization of $\cK$. The trace
\qq
\thol{\a}{\cK}=\Tr_{\a}\Pexp\left(\int_0^1 A(t)dt\right)
\label{f.5}
\qqq
can be considered as a partition function of the quantum theory on
the coadjoint orbit of $SU(2)$, $A(t)$ playing the role of an
external classical field. We will try to put the trace~(\ref{f.5}) in
an exponential form similar to that of eq.~(\ref{1.13}).

We start with a simple particular case of the Campbell-Hausdorf
formula:
\qq
e^v e^w=\exp\left[v+\sum_{n=0}^{\infty}(-1)^n
\frac{B_n}{n!}\left({\rm ad}_v\right)^n w +\cO(w^2)\right],
\label{f.6}
\qqq
here $v$ and $w$ are elements of a Lie algebra (say, $SU(2)$), ${\rm
ad}_v w=[v,w]$ and $B_n$ are Bernoulli numbers.

According to the Campbell-Hausdorf formula,
\qq
e^{v_1}\ldots e^{v_n}=\exp[\cv],
\label{f.7}
\qqq
here $\cv$ is a Lie algebra valued infinite polynomial in commutators
of $v_i$. We denote as $\cvn$ a part of $\cv$ containing the $n$th
order monomials which are linear in all $v_i$:
\qq
\cvna=\caa v_1^{a_1}\ldots v_n^{a_n}.
\label{f.8}
\qqq
A simple corollary of eq.~(\ref{f.6}) is the following
%%%%%%%%%%%%%%%%%%%%%%%%%%%%%%%%%%%%%%%%%%%%%%%%%%%%%%%
\begin{proposition}
\label{pf.1}
The coefficient $\caa$ is a sum of tensors coming from the diagrams
like those of Figs.~1-4 according to the rule: a dashed line with $m$
vertices which is depicted in Fig.~1, goes from
the upper right to the lower
left and produces a factor
\qq
(-1)^m\frac{B_m}{m!}\sum_{s\in S_m}f^a_{b,a_{s(1)},\ldots,a_{s(m)}},
\label{f.9}
\qqq
here $S_m$ is a group of permutations of numbers $1,\ldots,m$ and the
tensor $f^a_{b,a_1,\ldots,a_m}$ is a sum of products of structure
constants of the group $SU(2)$:
\qq
f^a_{b,a_1,\ldots,a_m}=(-2)^m\epsilon_{a_1 b b_1}
\epsilon_{a_2 b_1 b_2}\cdots \epsilon_{a_m b_{m-1} a},
\label{f.10}
\qqq
the sum over repeated indices is, of course, implied in
eq.~(\ref{f.10}). A contribution of the whole diagram is a sum of the
factors~(\ref{f.9}) of its dashed lines over the intermediate
indices.

More generally, if we pick $m$ elements $v_{i_1},\ldots,v_{i_m}$,
$1\leq i_1<\ldots i_m\leq n$, then the $m$th order homogeneous part of
$\cv$ which is linear in them, is equal to
\qq
C^a_{a_{i_1},\ldots,a_{i_m}}v^{a_{i_1}}_{i_1}\cdots
v_{i_m}^{a_{i_m}}.
\label{f.11}
\qqq
\end{proposition}
%%%%%%%%%%%%%%%%%%%%%%%%%%%%%%%%%%%%%%%%%%%%%%%%%%%%%%%%%%%%%%%%%%

\noindent
\underline{Examples}
\nopagebreak

The diagram for $n=2$ is drawn in Fig.~2:
\qq
i\sigma_aC^a_{a_1,a_2}v_1^{a_1}v_2^{a_2}=\frac{1}{2}[v_1,v_2].
\label{f.12}
\qqq
The diagrams for $n=3$ are drawn in Fig.~3:
\begin{eqnarray}
i\sigma_a C^a_{a_1,a_2,a_3}v_1^{a_1}v_2^{a_2}v_3^{a_3}&=&
\frac{1}{4}[[v_1,v_2],v_3]+\frac{1}{12}
\left([v_1,[v_2,v_3]]+[v_2,[v_1,v_3]]\right)
\nonumber\\
&=&\frac{1}{6}\left([v_1,v_2],v_3]+[v_1,[v_2,v_3]]\right).
\label{f.13}
\end{eqnarray}
The diagrams for $n=4$ are drawn in Fig.~4:
\begin{eqnarray}
i\sigma_aC^a_{a_1,a_2,a_3,a_4}&=&
\frac{1}{180}(
[v_1,[v_2,[v_3,v_4]]]+[v_1,[v_3,[v_2,v_4]]]
+[v_2,[v_1,[v_3,v_4]]]+[v_2,[v_3,[v_1,v_4]]]
\nonumber\\
\sshift{
+[v_3,[v_1,[v_2,v_4]]]+[v_3,[v_2,[v_1,v_4]]])
}
\nonumber\\
&&+\frac{1}{24}([[v_1,v_2],[v_3,v_4]]+[v_3,[[v_1,v_2],v_4]])
\nonumber\\
&&+\frac{1}{24}([[v_1,v_3],[v_2,v_4]]+[v_2,[[v_1,v_3],v_4]])
\nonumber\\
&&+\frac{1}{24}([[v_2,v_3],[v_1,v_4]]+[v_1,[[v_2,v_3],v_4]])
\nonumber\\
&&+\frac{1}{8}[[[v_1,v_2],v_3],v_4].
\label{f.14}
\end{eqnarray}

The Proposition~\ref{pf.1} allows us to ``take a logarithm'' of the
parallel transport operator $\Pexp\left(\int_0^1 A(t)dt\right)$. We
introduce an ``iterated'' integral (see also~\cite{Ch}):
\qq
\int_0^1 dt_1\ldots dt_n\{A(t_1),\ldots,A(t_n)\}=
\int_{0\leq t_n\leq\ldots\leq t_1\leq 1}
A(t_1)\ldots A(t_n).
\label{f.15}
\qqq
Note that the iterated integral depends on a choice of the zero point
of $t$ parametrization of $\cK$.
%%%%%%%%%%%%%%%%%%%%%%%%%%%%%%%%%%%%%%%%
\begin{proposition}
\label{pf.2}
The holonomy operator $\Pexp\left(\int_0^1 A(t)dt\right)$ can be
presented as an exponential of an infinite sum of iterated integrals:
\qq
\Pexp\left(\int_0^1 A(t)dt\right) =
\exp\left[i\sigma_a\sum_{n=1}^\infty
\caa\int_0^1dt_1\ldots dt_n\{A^{a_1}(t_1),\ldots,
A^{a_n}(t_n)\}\right].
\label{f.16}
\qqq
\end{proposition}
%%%%%%%%%%%%%%%%%%%%%%%%%%%%%%%%%%%%%%%%

Let us first present a simple ``physical'' proof of
eq.~(\ref{f.13}). We split the interval $0\leq t\leq 1$ into many
small intervals $\Delta t_i$ with middle points $t_i$ so that
\qq
\Pexp\left(\int_0^1 A(t)dt\right)=
e^{A(t_1)\Delta t_1} e^{A(t_2)\Delta t_2}\cdots
e^{A(t_n)\Delta t_n}.
\label{f.17}
\qqq
Then we apply the Campbell-Hausdorf formula to the r.h.s. of
eq.~(\ref{f.17}) retaining only those terms of $C(A(t_1)\Delta
t_1,\ldots,A(t_n)\Delta t_n)$ which are at most linear in any
particular $A(t_i)$. According to the Proposition~\ref{pf.1}, such
terms are given by eq.~(\ref{f.11}) with $v_{i_m}$ substituted by
$A(t_{i_m})$. It is
easy to see that a sum of all polynomials of a given
order converges to the iterated integral of eq.~(\ref{f.16}).

To prove eq.~(\ref{f.16}) more rigorously we may use the following
presentation of the holonomy operator:
\begin{eqnarray}
\Pexp\left(\int_0^1A(t)dt\right)&=&
\sum_{n=0}^\infty \int_0^1 dt_1\ldots dt_n
\{A(t_1),\ldots,A(t_n)\}
\nonumber\\
&=&\sum_{n=0}^\infty\frac{1}{n!}\sum_{s\in S_n}
\int_0^1 dt_1,\ldots,dt_n
\{A(t_{s(1)},\ldots,A(t_{s(n)})\}
\nonumber\\
&\stackrel{\rm def.}{=}&
\sum_{n=0}^\infty\frac{1}{n!}
\int_0^1 dt_1,\ldots,dt_n
{\rm P}[A(t_{1},\ldots,A(t_{n})],
\label{f.18}
\end{eqnarray}
here ${\rm P}[A(t_{1},\ldots,A(t_{n})]$ is a path-ordered
product, i.e. the Lie algebra valued forms $A(t_i)$ are multiplied in
the order of the values of their arguments $t_i$: for the largest
$t_i$, $A(t_i)$ stands to the left and so on.

The Proposition~\ref{pf.1} implies the following formula for the
product of $n$ Lie algebra elements $v_i$:
\qq
v_1\ldots v_n=\partial_{\epsilon_1}\ldots\partial_{\epsilon_n}\left.
\sum_{m=1}^n\frac{1}{m!}\left(i\sigma_a\sum_{l=1}^m\sum_{s\in S^l_n}
C^a_{a_1,\ldots,a_l}v_{s(1)}^{a_1}\ldots v_{s(l)}^{a_l}
\epsilon_{s(1)}\ldots \epsilon_{s(1)}\right)^m
\right|_{\epsilon_1=\ldots =\epsilon_n=0}\!\!\!,
\label{f.19}
\qqq
here $S_n^l$ is a set of all injections of $l$ numbers $1,\ldots, l$
into $n$ numbers $1,\ldots, n$ which keeps the order. A more explicit
version of this formula requires a splitting of $n$ elements $v_i$
into $m$ sets, with $n_i$ elements in each set, $m$ being an
arbitrary integer number: $1\leq m\leq n$. Denote by
$(s_1,\ldots,s_m)$ an injection of the union of $m$ sets
$1,\ldots,n_i$ into the set $1,\ldots,n$ which preserves the order
within each set $1,\ldots,n_i$: $s_l(i)>s_l(j)$ if $i>j$. Consider
now a symmetrized product
\qq
D=\frac{1}{m!\prod_{l=1}^m(\#l)!}
\sum_{s\in S_m}D_{s(1)}\ldots D_{s(m)},
\label{f.20}
\qqq
here
\qq
D_i=i\sigma_a C^a_{a_1,\ldots, a_{n_i}}
v^{a_1}_{s_i(1)}\ldots v^{a_{n_i}}_{s_i(n_i)}
\label{f.21}
\qqq
and $\#l$ is the number of indices $i$ for which $n_i=l$. The r.h.s.
of eq.~(\ref{f.19}) is equal to the sum of all such $D$ taken over
all numbers $m$, all possible splittings and all injections
$(s_1,\ldots,s_m)$.

Let us apply this presentation to the product $A(t_{s(1)})\ldots
A(t_{s(n)})$
appearing in the second line of eq.~(\ref{f.18}). Suppose
that we permute some of $A(t_{s(i)})$. A term $D$ coming from a
particular injection $(s_1,\ldots,s_m)$ still remains if this
permutation does not change the order within the $n_i$-element sets
into which the elements $A(t_{s(i)})$ are split by the injection.
Therefore we can combine the integrals over $0\leq
A(t_{s(1)})\leq\ldots\leq A(t_{s(1)})\leq 1$ which come with a
particular term $D$ into one integral over the regions $0\leq
A(t_{s(s_i(n_i))})\leq\ldots\leq A(t_{s(s_i(1))})\leq 1$ for $1\leq
i\leq m$. This operation leaves a subgroup of $S$ which permutes the
images $s_i(1),\ldots,s_i(n_i)$ for a given $i$. Its composition with
$S_n^l$ creates a redundant group $S$. The sum over its elements can
be removed by relabelling the integration variables
and adding an extra factor $n!$ which cancels the same factor in the
denominator of eq.~(\ref{f.18}). It is not hard to see that what
remains is the $m$th order term in the expansion of the exponential
in eq.~(\ref{f.16}). This completes the proof of the
Proposition~\ref{pf.2}.

A presentation of the holonomy operator as an exponential enables us
to use the
Weyl character formula for the calculation of its trace. For
an element $v$ of the Lie algebra $su(2)$
\qq
\Tr_\a e^v=\frac{\sin(\a|\vv|)}{\sin|\vv|}=
\frac{|\vv|}{\sin|\vv|}\int_{|\va|=\a}
\frac{d^2\va}{4\pi|\va|}e^{i\va\cdot\vv}.
\label{f.22}
\qqq
Combining this equation with eq.~(\ref{f.16}) we conclude that
\begin{eqnarray}
\thol{\a}{\cK}=
\frac{\sin(\a|\vv|)}{\sin|\vv|}=
\frac{|\vv|}{\sin|\vv|}\int_{|\va|=\frac{\a}{K}}
\frac{K}{4\pi}\frac{d^2\va}{|\va|}e^{iK\va\cdot\vv},
\label{f.23}\\
v^a=\sum_{n=1}^{\infty}\caa
\int_0^1 dt_1,\ldots dt_n\{A^{a_1}(t_1),\ldots,A^{a_n}(t_n)\}.
\label{f.24}
\end{eqnarray}
This is the exponential presentation of the trace of holonomy that we
were looking for.

We can apply the formula~(\ref{f.23}) to the holonomies along the
link components of eq.~(\ref{1.1}):
\begin{eqnarray}
\label{f.25}
\lefteqn{
Z_{\a_1,\ldots,\a_n}(M,\cL;k)=
\spint\int[\cD A_\mu]\left(\prod_{j=1}^n
\frac{|\vv_j|}{\sin|\vv_j|}\right)
}
\\
&&
\times
\exp\left[-\frac{iK}{\pi}\epsilon^{\mu\nu\rho}\int_M d^3 x\left(
\frac{1}{2}\vA_\mu\cdot\partial_\nu\vA_\rho+
\frac{1}{3}\vA_\mu\cdot(\vA_\nu\times\vA_\rho)\right)+
iK\sum_{j=1}^n \va_j\cdot\vv_j\right],
\nonumber
\end{eqnarray}
here
\qq
v^a_j=\sum_{n=1}^\infty \caa\int_0^1
dt^{(j)}_1\ldots dt^{(j)}_n
\{A(t^{(j)}_1),\ldots,A(t^{(j)}_n)\},
\label{f.26}
\qqq
and $t^{(j)}$ is a parametrization of a link component $\cL_j$.
We put $K$ instead of $k$ as a factor in front of the integral in the
exponent of eq.~(\ref{f.25}) in order to be able to ignore the 1-loop
corrections to the propagator of the original Chern-Simons
theory~(\ref{1.1}) (see e.g.~\cite{BN} and references therein).
The formula~(\ref{f.25})
indicates that the terms $\va_j\cdot\vv_j$ may be
considered as extra vertices in the quantum Chern-Simons theory. In
other words, Feynman rules for the quantum theory~(\ref{f.25})
include a propagator (i.e., a Green's function)
\qq
\langle A_\mu^a(x_1) A_\nu^b(x_2)\rangle=
-\frac{i\pi}{K}\delta^{ab}\Omega_{\mu\nu}(x_1,x_2),
\label{f.27}
\qqq
a usual cubic vertex
\qq
V_3=-\frac{iK}{3\pi}\epsilon^{\mu\nu\rho}
\int_M d^3 x \vA_\mu\cdot(\vA_\nu\times\vA_\rho)
\label{f.28}
\qqq
and extra vertices coming from the holonomies along the link
components
\qq
V_n^{(j)}=iK\va_j\cdot \vC_{a_1,\ldots, a_n}
\int_0^1
dt^{(j)}_1\ldots dt^{(j)}_n
\{A(t^{(j)}_1),\ldots,A(t^{(j)}_n)\}.
\label{f.29}
\qqq
In particular,
\begin{eqnarray}
V_1^{(j)}&=&iK\va_j\cdot\left(\int_0^1
dt^{(j)} \vA(t^{(j)})\right),
\label{f.1029}\\
V_2^{(j)}&=&-iK\va_j\cdot \left(\int_0^1 dt_1^{(j)} dt_2^{(j)}
\{\vA(t_1^{(j)})
\stackrel{\displaystyle\times}{,}\vA(t_2^{(j)})\}\right),
\label{f.2029}\\
V_3^{(j)}&=&\frac{2}{3}iK\va_j\cdot\left[
\int_0^1 dt_1^{(j)}dt_2^{(j)} dt_3^{(j)}
\left(\{(\vA(t^{(j)}_1)\stackrel{\displaystyle\times}{,}
\vA(t^{(j)}_2))\stackrel{\displaystyle\times}{,}
\vA(t^{(j)}_3)\}
\right.\right.
\label{f.3029}\\
\sshift{
\left.\left.
+
\{\vA(t^{(j)}_1)\stackrel{\displaystyle\times}{,}
(\vA(t^{(j)}_2)\stackrel{\displaystyle\times}{,}
\vA(t^{(j)}_3))\}\right)\right],
}
\nonumber
\end{eqnarray}
here
\qq
\va_j\cdot\{\vA(t_1^{(j)})
\stackrel{\displaystyle\times}{,}\vA(t_2^{(j)})\}
=\epsilon_{abc}a_j^a\{A^b(t_1^{(j)}),A^c(t^{(j)}_2)\}.
\label{f.4029}
\qqq
A symmetric bilocal (1,1)-form (i.e., a 1-form in both variables
$x,y$) $\Omega_{\mu\nu}(x,y)$ of eq.~(\ref{f.27}) is a Green's
function of the operator $\epsilon^{\mu\nu\rho}\partial_\nu$ (i.e.,
of a differential $d$). It should satisfy an equation
\qq
d_y\Omega(x,y)=\delta^{(3)}(y-x) + d_x\tilde{\Omega}(x,y),
\label{f.30}
\qqq
here $\delta^{(3)}(y-x)$  is a 3-form $\delta$-function while
$\tilde{\Omega}(x,y)$ is a (0,2)-form, i.e. a 2-form in $y$ and a
0-form in $x$. If $M=S^3$ and $S^3$ is presented as $\IR^3$ with an
infinite point, then
\qq
\Omega_{\mu\nu}(x,y)=\frac{1}{4\pi}\epsilon_{\mu\nu\rho}
\frac{y^\rho-x^\rho}{|y-x|^3}.
\label{f.31}
\qqq
For more information on $\Omega(x,y)$ see, e.g.~\cite{BN} and
references therein.

There are some extra vertices coming from the expansion of the
prefactors $\frac{|\vv_j|}{\sin|\vv_j|}$ into the powers of
$\vv_j^2$. These vertices do not assemble into an exponential.
According to the standard combinatorics of Feynman diagrams, the
trivial connection contribution to the
path integral over $[\cD A_\mu]$ in eq.~(\ref{f.25}) can be
presented as a product of two factors:
\qq
\ztraml=\left(1+G(\va_1,\ldots,\va_n;K)\right)
\exp\left[\sum_{l=0}^\infty
K^{1-l} L^{(l)}\adots\right],
\label{f.32}
\qqq
here $G(\va_1,\ldots,\va_n;K)$ is a sum of all Feynman diagrams
containing at least one vertex coming from
$\frac{|\vv_j|}{\sin|\vv_j|}$ and $L^{(l)}\adots$ is a sum of all
connected  $l$-loop Feynman diagrams which contain only the
vertices~(\ref{f.28}) and~(\ref{f.29}). Each of these vertices carries
a factor $K$ while the propagator~(\ref{f.27}) is of order
$K^{-1}$. As a result, $l$-loop diagrams have a factor $K^{1-l}$ which
we made explicit in the exponent of eq.~(\ref{f.32}).  The vertices
coming from $\frac{|\vv_j|}{\sin|\vv_j|}$ do not carry the factor
$K$, therefore the diagrams that contain such vertices have only zero
or negative powers of $K$.  Thus, combining the Taylor series
expansions
\begin{eqnarray}
L^{(1)}\adots&=&\sum_{m=2}^\infty\lm,
\label{f.33}
\qqq
\qq
\label{f.34}
[1+G(\va_1,\ldots,\va_n;K)]\exp\left[\sum_{l=1}^\infty
K^{1-l} L^{(l)}\adots\right]
&&\hspace{-0.5in}\\
&&=1+\spml,
\nonumber
\end{eqnarray}
$\lm$ and $P_{l,m}\adots$
being invariant homogeneous polynomials of order
$m$, with eqs.~(\ref{f.32}) and ~(\ref{f.25}) we arrive at
Reshetikhin's formula~(\ref{5.1}).

A quadratic polynomial $L_2$ comes from the Feynman diagram
containing only one propagator~(\ref{f.27}) both endpoints of which
are attached to link components. Since
\qq
\oint_{\cL_i}dx_i\oint_{\cL_j}dx_j\Omega(x_i,x_j)=l_{ij},
\label{f.35}
\qqq
$l_{ij}$ being the gaussian linking number, we conclude that
\qq
L_2\adots=\sum_{i,j=1}^{n}l_{ij}\,\va_i\cdot\va_j.
\label{f.36}
\qqq
This completes the proof of the Proposition~\ref{pf2.1}.

We will also need in the future the polynomials
\begin{eqnarray}
L_3\adots&=&
\sum_{i,j,k=1}^{n}\lti\;\va_i\cdot(\va_j\times\va_k),
\label{f.37}\\
L_4\adots&=&\sum_{i,j,k,l=1}^{n}\lfi\;
(\va_i\times\va_j)\cdot(\va_k\times\va_l).
\label{f.38}
\end{eqnarray}
It will become clear from the study of Feynman diagrams in
Sections~\ref{*4} and~\ref{*5} why we use these particular group
weight structures.

Consider now the group weights associated to connected tree level
diagrams contributing to the polynomials $\lm$. These diagrams are
combinations of tree diagrams of the original Chern-Simons
theory~(\ref{1.2}) sewn by the new vertices~(\ref{f.29}) which are
themselves tree diagrams (see Figs.~1--4). In both types of tree
diagrams the segments are $\delta$-symbols $\delta_{ab}$ and the
elementary cubic vertices are proportional to the group structure
constants $\epsilon_{abc}$.

We may associate an invariant monomial
$\finv$ which is linear in all vectors $\vb_j$,
$1\leq j\leq m$ to a combined group weight coming from a Feynman
diagram with $m$ vertices~(\ref{f.29}) by placing vectors $\vb_j$ at
the bottom of the tree diagrams associated with its
vertices~(\ref{f.29}). Then in order to get the actual group weight
of the Feynman diagram that would contribute to the polynomial
$\lm$, we should substitute $n$ vectors $\va_j$ for $m$ vectors
$\vb_j$ depending on which vertex~(\ref{f.29}) comes from which link
component $\cL_j$. Since every  tree diagram with more than three
external legs contains at least two $Y$-shaped configurations with
two vectors $\vb$ attached to each of them (see Fig.~5)
and since the cubic
vertices produce antisymmetric tensors $\epsilon_{abc}$, we conclude
that a polynomial $\finv$, $m\geq 4$ is zero if at
least $m-1$ of $m$ vectors $\vb_j$ are parallel. It is obvious that
the same is true for $m=3$.
%
%%%%%%%%%%%%%%%%%%%%%%%%
\begin{proposition}
\label{pf.3}
The polynomials $\lm$ are produced from invariant homogeneous
polynomials $\finv$ of order $m$ by substituting $n$ vectors $\va_j$
in place of $m$ vectors $\vb_j$. The polynomials $\finv$, $m\geq 3$
are equal to
zero if at least $m-1$ of $m$ vectors $\vb_j$ are parallel.
\end{proposition}
%%%%%%%%%%%%%%%%%%%%%%%%%
%
This proposition will play an important role in extracting the
multivariable Alexander polynomial from the r.h.s. of
eq.~(\ref{5.1}) in~\cite{RoII}.

Suppose for a moment that the link $\cL$ has only one component, i.e.
it is in fact a knot $\cK$. An immediate consequence of the
Proposition~\ref{pf.3} is that only a quadratic term survives in the
exponent of the Reshetikhin's formula~(\ref{5.1}). As a result, the
whole formula is reduced to eq.~(\ref{2.15}). Thus we produced yet
another proof of the first part of the Melvin-Morton
conjecture~\ref{1.1}.

%*****************************************
\nsection{The Exponent of Reshetikhin's Formula}
\label{*3}
%*****************************************

In the remainder of this paper we are going to study more closely the
structure of the exponent of Reshetikhin's formula~(\ref{5.1}).
%%%%%%%%%%%%%%%%%%%%%%%%%%%%%%%%%%
\begin{conjecture}
\label{cf3.1}
If $L_l\adots=0$ for all $l<m$, then the coefficients of the
polynomial $\lm$ are proportional to the $m$th order Milnor's linking
numbers $\lmi$ of the link $\cL$:
\qq
\lm=\frac{(i\pi)^{m-2}}{m}
\sum_{1\leq i_1,\ldots,i_m\leq n}
\lmi\Tr (\vgs\cdot\va_{i_i})
\cdots(\vgs\cdot\va_{i_m}),
\label{n.1}
\qqq
here $\vgs=(\sigma_1,\sigma_2,\sigma_3)$ is a 3-dimensional vector
formed by Pauli matrices. \end{conjecture}
%%%%%%%%%%%%%%%%%%%%%%%%%%%%%%%%%%
We use the scalar product $\vgs\cdot\va_i$
instead of simply $a_i$ in order to stress that we are dealing with
the $su(2)$ Lie algebra element in the fundamental representation.
Note that $l^{(\mu)}_{ij}=l_{ij}$, so eq.~(\ref{n.1}) for $m=2$ is
consistent with eq.~(\ref{5.4}).

It is known that Milnor's link invariants of
the higher order are not well defined (at least, in $\ZZ$) if lower
order invariants are non-zero. The same property is shared by the
polynomials $L_m$. Their coefficients can not be restored
unambiguously from the value of the partition function $\ztraml$
because there are changes in integration variables $\va_j$ which keep
the form of eq.~(\ref{5.1}) but still alter the coefficients of
higher order polynomials $L_m$ if the lower order polynomials are
non-zero.
Suppose for example that
we substitute a vector $\va_j$ by a vector $\va_j^\prime$
obtained by rotating $\va_j$ around another vector $\va_k$ by an angle
$\varphi_{j,k}$:
\qq
\va_j^\prime=\sum_{m=0}^{\infty}
\frac{\varphi_{j,k}^m}{m!}
\underbrace{
[\va_k\times[\va_k\times\ldots[\va_k\times\va_j]\ldots]
}_{m\;\;\;{\rm times}}.
\label{sp.1}
\qqq
A substitution of eq.~(\ref{sp.1}) into the quadratic polynomial $L_2$
of eq.~(\ref{f.36}) generates, among others, new cubic terms, so that
\qq
l^{\prime(3)}_{ijk}=\lti+\l_{ij}\varphi_{j,k},
\;\;\;\;\;\;1\leq i\leq n,\;\;\;\;\;\;i\neq j,k.
\label{sp.2}
\qqq
To put it differently, we need to know the coefficients $\lti$ only up
to these transformations. In fact, as we will see in
Section~\ref{*4}, the Feynman rules of Section~\ref{*2} predict
these coefficients only up to this transformation due to the
dependence of the vertices~(\ref{f.29}) on the choice of a zero point
in $t^{(j)}$ parametrization of the link component $\cL_j$. However
this ambiguity disappears if the gaussian linking numbers $l_{ij}$
are equal to zero. Then the coefficients $\lti$ are well defined and
turn out to be proportional to the triple Milnor's invariants.

There are many other possible rotations of the integration variables
$\va_j$. For example, a vector $\va_j$  can be rotated around the
vector $\va_i\times\va_j$. This transformation will change the
coefficient
$l^{(4)}_{ij,ij}$ by an amount proportional  to $l_{ij}$. It will
also cause a change in the coefficient $p_{ij}$ of the
preexponential polynomial
\qq
P_{0,2}\adots=\sum_{i,j=1}^{n}p_{ij}\va_i\cdot\va_j.
\label{5.3}
\qqq
due to a nontrivial jacobian factor.

Let us now consider some circumstantial evidence in support of the
Conjecture~\ref{cf3.1}. Milnor's linking numbers are invariant under
tying a small knot on a link component.
Let $\cL$ be an
$n$-component knot in a \rhs
$M$ and $\cK$ be a knot in $S^3$. We cut an infinitely
small 3-dimensional ball $B^3$, whose center belongs to $\cL_1$,
out of $M$ and we cut another infinitely small
ball $\tilde{B}^3$, whose center
belongs to $\cK$, out of $S^3$. Then we glue the boundaries
$\partial(M\setminus B^3)$ and $\partial(S^3\setminus\tilde{B}^3)$,
thus producing a new link $\cL\p$ in $M$. In other words, we ``tie''
a small knot $\cK$ on the component $\cL_1$ of $\cL$. Milnor's
invariants of $\cL$ and $\cL\p$ coincide.

Let us see what happens to the exponent of eq.~(\ref{5.1}). According
to ~\cite{Wi1},
\qq
Z^{({\rm tr})}_{\a_1,\ldots,\a_n}(M,\cL\p;k)=
Z^{({\rm tr})}_{\a_1,\ldots,\a_n}(M,\cL;k)
Z_{\a_1}(S^3,\cK;k)
\sqrt{\frac{K}{2}}
\frac{1}{\sin\left(\frac{\pi}{K}\a_1\right)}.
\label{sp4.1}
\qqq
Combining the formula~(\ref{5.1}) for $\ztraml$
%$Z^{({\rm tr})}_{\a_1,\ldots,\a_n}(M,\cL\p;k)$
with the
formula~(\ref{2.15}) for
$Z_{\a_1}(S^3,\cK;k)$
we can  easily derive
the formula~(\ref{5.1}) for
$Z^{({\rm tr})}_{\a_1,\ldots,\a_n}(M,\cL\p;k)$:
\begin{eqnarray}
\lefteqn{
Z^{({\rm tr})}_{\a_1,\ldots,\a_n}(M,\cL\p;k)=
}
\label{sp4.2}\\
&&
=\ztrmk\spint\exp\left[\frac{i\pi K}{2}\left(
\nu\va_1^2+\sum_{m=2}^{\infty}\lm\right)\right]
\nonumber\\
\shift{
\times\onespml
e^{-\frac{i\pi}{2K}\nu}
\frac{\sin\left(\frac{\pi}{K}\right)}
{\left(\frac{\pi}{K}\right)}
\frac{\pi |\va_1|}{\sin(\pi |\va_1|)}
J(K|\va_1|,K).
}
\nonumber
\end{eqnarray}
As we see, the exponent remains the same except for the trivial
change of framing of $\cL_1$: $l_{11}\p=l_{11}+\nu$.
This provides a confirmation for
our conjecture that the coefficients of the polynomials $\lm$ are
proportional to Milnor's linking numbers.

%***************************************
\nsection{Flat Connections in the Link Complement}
\label{*n1}
%***************************************

The strongest evidence in support of the Conjecture~\ref{cf3.1} is
provided by the relation between the large $K$  asymptotics of the
Jones polynomial of a link and the flat connections in the link
complement. In this section we will assume for simplicity that our
manifold $M$ is a 3-dimensional sphere $S^3$. Consider a large $K$
limit of the Jones polynomial~(\ref{1.1}) when the ratios~(\ref{5.2})
are kept constant. Then (see~\cite{RoI} and references therein) the
invariant $\zasl$ can be expressed as a path integral over the
connections in the link complement:
\begin{eqnarray}
\zasl&=&\int [\cD A_\mu]\exp\left(\frac{i}{\hbar}
S_{CS}\p[A_\mu]\right),
\label{n1.1}\\
\label{n1.2}
S\p_{CS}[A_\mu]&=&
\frac{1}{2}\Tr \epsilon^{\mu\nu\rho}
\int_{S^3\setminus\sum_{j=1}^n\Tub(\cL_j)}
d^3 x\left(A_\mu\partial_\nu A_\rho -
\frac{2}{3}A_\mu A_\nu A_\rho\right)
\\
\sshift{
-\frac{1}{2}\sum_{j=1}^n \Tr\left[
\left(\oint_{C_1^{(j)}}A_\mu dx^\mu\right)
\left(\oint_{C_2^{(j)}}A_\mu dx^\mu\right)\right].
}
\nonumber
\end{eqnarray}
In this formula $C^{(j)}_1$ and $C^{(j)}_2$ are two basic cycles on
the boundary of the tubular neighborhood $\Tub(\cL_j)$. The cycle
$C^{(j)}_1$ is a meridian of $\cL_j$, it can be contracted through
$\Tub(\cL_j)$. A cycle $C_2^{(j)}$ is a parallel, it has a unit
intersection number with $C^{(j)}_1$ and it is defined only modulo
$C^{(j)}_1$. A self-linking number $l_{jj}$ is by definition a
linking number between $\cL_j$ and $C^{(j)}_2$. The path
integral~(\ref{n1.1}) goes over the gauge equivalence classes of
connections $A_\mu$ which satisfy the boundary conditions
\qq
\hol{C_1^{(j)}}
=\exp\left(\frac{i\pi}{K}\sigma_3\a_j\right)\equiv
\exp(i\pi\vgs\cdot\va_j)\;\;\;\;\;\;
{\rm up\;\;\; to\;\;\; a\;\;\; conjugation}.
\label{n1.3}
\qqq
The large $K$ limit of the path integral~(\ref{n1.1}) can be found
with the help of the stationary phase  approximation. The invariant
$\zasl$ will be presented in the form~(\ref{1.13}), however this time
the sum will go over the flat connections in the link complement
$S^3\setminus\sum_{j=1}^n\Tub(\cL_j)$ satisfying the boundary
conditions~(\ref{n1.3}). On the other hand, the same large $K$
asymptotics of $\ztrasl$ can be found  by applying the stationary
phase approximation to the finite dimensional integral over the
vectors $\va_j$ in Reshetikhin's formula~(\ref{5.1}). The invariant
will be presented as a sum over the conditional stationary
points of the phase
\qq
\sum_{m=1}^\infty\lm,
\label{n1.4}
\qqq
the conditions being eqs.~(\ref{5.2}). Therefore we conjecture that
there is a one-to-one correspondence between the flat connections in
the link complement, which are close to the trivial connection, and
the stationary phase points of the phase~(\ref{n1.4}), so that their
contributions to $\ztrasl$ are equal.

Let us make this relation more precise. Consider a 1-parametric
family of flat connections $A_\mu(x,\tau)$, $\tau\geq 0$ in the link
complement, which starts at the trivial connection: $A_\mu(x,0)=0$.
The holonomies of a flat connection define a homomorphism from the
fundamental group of the link $\pi_1(\cL)$ into $SU(2)$. In
Wirtinger's presentation of $\pi_1(\cL)$ the link is projected onto a
plane and the fundamental group is generated by the meridians
$\cC_{j,i}$ of the pieces into which the link components $\cL_j$ are
split by overcrossings. If two pieces $\cL_{j,i}$ and $\cL_{j,i+1}$
of a link component $\cL_j$ are joined at the overcrossing of $\cL_j$
by a piece $\cL_{k,l}$ of $\cL_k$, then the elements
$\cC_{j,i},\cC_{j,i+1},\cC_{k,l}\in\pi_1(\cL)$ satisfy the equation
\qq
\cC_{j,i+1}=\cC_{k,l}^{\pm 1}\cC_{j,i}\cC_{k,l}^{\mp 1},
\label{n1.5}
\qqq
the signs in the exponents depend on the signature of the
overcrossing. These relations imply that for a given $j$ the
holonomies of $A_\mu(x,\tau)$ along the meridians $\cC_{j,i}$ are
all equal to the leading order in $\tau$:
\qq
\Pexp\left(\oint_{\cC_{j,i}}A_\mu(x,\tau)dx^\mu\right)
=\exp[i\pi\tau\vgs\cdot\va\p_j+\cO(\tau^2)],
\label{n1.6}
\qqq
here $\va\p_j$ are the vectors indicating the directions in the Lie
algebra $su(2)$ in which the trivial connection of $\tau=0$ in
deformed as $\tau$ grows.

Consider a knot $\cK$ in the link complement
$S^3\setminus\sum_{j=1}^n\Tub(\cL_j)$, we denote as $l_{0j}$ the
linking numbers of $\cK$ and $\cL_j$. It is not hard to see that to
the leading order in $\tau$
\qq
\Pexp\left(\oint_{\cK}A_\mu(x,\tau)dx^\mu\right)=
\exp\left[i\pi\tau\vgs\cdot\left(
\sum_{j=1}^nl_{0j}\va\p_j\right)+\cO(\tau^2)\right],
\label{n1.7}
\qqq
so that if we attach a $\b$-dimensional representations to $\cK$,
then
\qq
\Tr_\b\Pexp\left(\oint_{\cK}A_\mu(x,\tau)dx^\mu\right)=
\beta\left[1-\tau^2\frac{\pi^2}{6}(\b^2-1)
\left(\sum_{j=1}^nl_{0j}\va\p_j\right)^2\right]+\cO(\tau^3).
\label{n1.8}
\qqq
Consider now a family of conditional stationary points of the
phase~(\ref{n1.4}):
\qq
\va_j^{({\rm st})}(\tau)=
\va\pp_j\tau+\cO(\tau^2).
\label{n1.9}
\qqq
Let us add the knot $\cK$ as a 0th component to the link $\cL$ (i.e.,
multiply the r.h.s. of eq.~(\ref{1.1}) by $\thol{\b}{\cK}$)
and see what happens to the contribution of the stationary
phase points~(\ref{n1.9}). The new exponent of the
formula~(\ref{5.1}) should include the terms containing the vector
$\vb$ corresponding to $\cK$: $|\vb|=\b/K$. We assume that $\b\sim
1$ as $K\rightarrow \infty$. Then, to the leading order in $\tau,K$
and $\b$, we should account only for the bilinear term
$\vb\cdot\left(\sum_{j=1}^nl_{0j}\va_j\right)$ in the new exponent.
As a result, the contribution of the stationary phase
point~(\ref{n1.9}) is multiplied by the factor
\qq
\int_{|\vb|=\frac{\b}{K}}
\frac{K}{4\pi}\frac{d^2\vb}{|\vb|}
\exp\left[i\pi\tau K\vb\cdot \left(\sum_{j=1}^n
l_{0j}\va_j\pp\right)\right]
=
\b\left[1-\frac{1}{6}\pi^2\tau^2\b^2
\left(\sum_{j=1}^n l_{0j}\va_j\pp\right)^2\right]
+\cO(\tau^3).
\label{n1.10}
\qqq
This factor should be interpreted as the trace of the holonomy of the
flat connection corresponding to the stationary point~(\ref{n1.9}).
Comparing eqs.~(\ref{n1.10}) and~(\ref{n1.8}) we conclude that
\qq
\va_j\p=\va_j\pp
\label{n1.11}
\qqq
for the family of the flat connections $A_\mu(x,\tau)$ that
corresponds to the family of the stationary phase
points~(\ref{n1.9}).

Milnor's linking numbers $\lmi$ allow us to formulate the necessary
conditions that the vectors $\va_j\p$ have to satisfy so that the
flat connections $A_\mu(x,\tau)$ with the holonomies~(\ref{n1.6})
exist. Let us briefly review the algebraic definition of these
numbers. Consider a parallel $\tilde{\cC}_j$, by definition it is an
element of $\pi_1(\cL)$ which is homologically equivalent to the
parallel $C^{(j)}_2$ and also commutes with the meridian $\cC_{j,1}$.
The element $\tilde{\cC}_j$ can be expressed as a product of powers
of the meridians $\cC_{k,l}$ since they generate the whole group
$\pi_1(\cL)$. Milnor showed~\cite{Mi} that the parallel
$\tilde{\cC}_j$ can be expressed only in terms of the meridians
$\cC_{k,1}$ (one meridian per link component) modulo the elements of
the lower central subgroup $\pi_1^{(q)}(\cL)$ of $\pi_1(\cL)$
($\pi_1^{(1)}(\cL)=\pi_1(\cL),\;\;
\pi_1^{(n+1)}(\cL)=[\pi_1^{(n)},\pi_1(\cL)]$) for any arbitrarily big
value of $q\in\ZZ$. Suppose that the first non-zero Milnor's numbers
appear at order $m$. Then we choose $q>m$. It is easy to see that for
the family of flat connections~(\ref{n1.6}) the holonomy along the
elements of $\pi_1^{(q)}$ is equal to 1 up to the order $\tau^q$. We
are interested in the holonomy along $\tilde{\cC}_j$ up the order
$\tau^{m-1}$, so we can neglect the elements of $\pi_1^{(q)}(\cL)$
and use the expression for $\tilde{\cC}_j$ in terms of $\cC_{k,1}$
modulo $\pi_1^{(q)}(\cL)$. Then, according to eqs.~(\ref{n1.6}) and
Milnor's definition of $\lmi$ as the coefficients of the Magnus
expansion of this expression,
\begin{eqnarray}
\lefteqn{
\Pexp\left(\oint_{\tilde{\cC}_j}A_\mu(x,\tau)dx^\mu\right)
}
\hspace*{2.5cm}
\label{n1.12}\\
&&
=1+(i\pi\tau)^{m-1}
\sum_{1\leq i_1,\ldots,i_{m-1}\leq n}
\lmij (\vgs\cdot\va_{i_1}\p)\cdots(\vgs\cdot\va_{i_{m-1}}\p)
+\cO(\tau^m)
\nonumber
\end{eqnarray}
(we actually used the relation $l_{i,1}=\exp[i\pi\tau\vgs\cdot\va_i]$
coming from eq.~(\ref{n1.6}) rather than $\cC_{i,1}=1+X_i$, $X_j$
being an indeterminate, which is a standard form of the Magnus
expansion, see also ~\cite{BN2}). Since the parallel $\tilde{\cC}_j$
commutes with the meridian $\cC_{j,1}$, we come to the following
conclusion:
%%%%%%%%%%%%%%%%%%%%%%%%%%%%%%%%%%%%%%%%%%%%%
\begin{proposition}
\label{pn1.1}
If the 1-parametric family of flat connections $A_\mu(x,\tau)$
defined by eq.~(\ref{n1.6}) exists in the link complement, then the
vectors $\va_j\p$ satisfy the condition
\qq
\left[\vgs\cdot\va_j\p,
\sum_{1\leq i_1,\ldots,i_{m-1}\leq n}
\lmij (\vgs\cdot\va_{i_1}\p)\cdots(\vgs\cdot\va_{i_{m-1}}\p)
\right]=0,\;\;\;\;1\leq j\leq n.
\label{n1.13}
\qqq
\end{proposition}
%%%%%%%%%%%%%%%%%%%%%%%%%%%%%%%%%%%%%%%%%%%%%%%%%%%%

Let us compare eq.~(\ref{n1.13}) with the stationary phase condition
for the integral of eq.~(\ref{5.1}). Suppose that $L_l\adots=0$ for
$l<m$. Then eq.~(\ref{n1.9}) presents a family of conditional
stationary points of the phase~(\ref{n1.4}) if
\qq
\frac{\partial L_m(\va_1\pp,\ldots,\va_n\pp)}
{\partial\va_j\pp}
\times\va_j\pp=0.
\label{n1.14}
\qqq
Then it follows from the invariance of Milnor's linking numbers
$\lmi$ under a cyclic permutation of the indices that
%%%%%%%%%%%%%%%%%%%%%%%%%%%%%%%%%%%%%%%%%%%%%%%%%
\begin{proposition}
\label{pn1.2}
The identification~(\ref{n1.11}) of the flat connections~(\ref{n1.6})
with conditional stationary points~(\ref{n1.9}) of the
phase~(\ref{n1.4}) is consistent with the conjectured
expression~(\ref{n.1}) for the polynomials $\lm$.
\end{proposition}
%%%%%%%%%%%%%%%%%%%%%%%%%%%%%%%%%%%%%%%%%%%%%%%%%

This proposition supports the Conjecture~\ref{cf3.1} but it does not
help us to fix the coefficient in front the trace in eq.~(\ref{n.1}).
This can be done by comparing the dominant part of the
stationary phase  of eq.~(\ref{5.1})
\qq
\frac{i\pi K}{2}\tau^m L_m(\va_1\p,\ldots,\va_n\p)
\label{n1.15}
\qqq
with the Chern-Simons action~(\ref{n1.2}) of the flat connection
$A_\mu(x,\tau)$. The action $S\p_{CS}$ has the following property
(see, e.g.~\cite{KiKl}): if the connections $A_\mu$ and
$A_\mu+\delta A_\mu$ are both flat, then
\qq
S_{CS}\p[A_\mu+\delta A_\mu]-S_{CS}\p[A_\mu]=
-\sum_{j=1}^n\Tr\left[\left(\oint_{C_1^{(j)}}
\delta A_\mu dx^\mu\right)\left(\oint_{C_2^{(j)}}A_\mu
dx^\mu\right)\right]
\label{n1.16}
\qqq
(this is a general property of a classical action:
$\delta\left.\left(\int_{t_1}^{t_2}[p\dot{q}-H(p,q)]dt\right)=
p\delta q\right|_{t_1}^{t_2}$). On the other hand, eq.~(\ref{n1.12})
implies that
\qq
\oint_{C_2^{(j)}} A_\mu(x,\tau)dx^\mu=
(i\pi\tau)^{m-1}
\sum_{1\leq i_1,\ldots,i_{m-1}\leq n}
\lmij (\vgs\cdot\va_{i_1}\p)\cdots(\vgs\cdot\va_{i_{m-1}}\p)
+\cO(\tau^m).
\label{n1.17}
\qqq
Combining eqs.~(\ref{n1.16}) and~(\ref{n1.17}) we conclude that
\qq
\frac{dS_{CS}\p[A_\mu(x,\tau)]}{d\tau}=
-(i\pi)^m\tau^{m-1}
\Tr
\sum_{1\leq i_1,\ldots,i_m\leq n}
\lmi (\vgs\cdot\va_{i_1}\p)\cdots(\vgs\cdot\va_{i_m}\p).
\label{n1.18}
\qqq
After integrating this equation over $\tau$ we arrive at
eq.~(\ref{n.1}).

%****************************************
\nsection{A Triple Milnor's Linking Number}
\label{*4}
%****************************************
Milnor's invariants $\lmi$ can be expressed as integrals of
differential forms constructed with the help of the Massey product
(see ~\cite{Po},~\cite{Tu} and references therein, a simple
introduction into this subject together with the relevant formulas
can be found in~\cite{MoRe}). We are going to check the
Conjecture~\ref{cf3.1} for the polynomials $L_3$ and $L_4$  by
comparing the Feynman diagram formulas for their coefficients with
these expressions.

\noindent
\underline{Preliminaries}
\nopagebreak

We start by introducing some useful notations. Let $\cL$ be an
$n$-component link in a \rhs $M$. Suppose that we cut out a tubular
neighborhood $\Tub(\cL_j)$ from the
manifold $M$ and then glue it back after switching its meridian
$C^{(j)}_1$ and its parallel $C^{(j)}_2$. We call such procedure an
$S$-surgery on $\cL_j$. With a slight abuse of notations, we denote
as $\Tub\p(\cL_j)$ the tubular neighborhood when it is glued back as
a result of $S$-surgery. We denote as $M_{i_1\ldots
i_m,\bar{j}_1\ldots \bar{j}_l}$ the manifold constructed from $M$ by
removing the tubular neighborhoods
$\Tub(\cL_{i_1}),\ldots,\Tub(\cL_{i_m})$ (we will also assume these
neighborhoods to be infinitely thin) and performing $S$-surgeries on
the link components $\cL_{j_1},\ldots\cL_{j_l}$.

We denote as $\omega_j$ a closed 1-form defined in $M_j$ by a
condition
\qq
\oint_{C_1^{(j)}}\omega_j=1
\label{sp.3}
\qqq
This form can be expressed with the help of the Green's
function~(\ref{f.27}):
\qq
\omega_j(\cdot)=\oint_{\cL_j}\Omega(t^{(j)},\cdot),
\label{sp1.18}
\qqq
here $t^{(j)})$
is a parametrization of $\cL_j$ and we slightly abused
the notations by using $t^{(j)}$ instead of $x(t^{(j)}$ as the
argument of $\Omega$. The linking numbers $l_{ij}$ can be expressed
with the help of the forms $\omega_j$:
\qq
l_{ij}=\oint_{\cL_i}\omega_j=\oint_{\cL_j}\omega_i
=\oint_{\cL_i}\oint_{\cL_j}\Omega(t^{(i)},t^{(j)}).
\label{sp.5}
\qqq
Another useful property of $\omega_j$ is that if $\omega$ is a smooth
1-form in an infinitely thin tubular neighborhood $\Tub(\cL_j)$, then
\qq
\int_{\partial\Tub(\cL_j)}
\omega_j\wedge\omega=\oint_{\cL_j}\omega.
\label{sp3.1}
\qqq

The following object appears naturally in the formulas for Milnor's
invariants. Let $\omega_1,\omega_2$ be two 1-forms defined on a knot
$\cK$ parametrized by $0\leq t\leq 1$. An ``iterated commutator''
$[\omega_1,\omega_2]$ is a bilocal (1,1)-form
\qq
[\omega_1,\omega_2](t_1,t_2)=\sign{t_1-t_2}
\omega_1(t_1)\omega_2(t_2).
\label{sp3.2}
\qqq
If both forms $\omega_1$ and $\omega_2$ are multilocal, then in our
notations
\begin{eqnarray}
\int_{\cK}[\omega_1,\omega_2]&\equiv&
\int_{\min(t_1,\ldots,t_m)>\max(t_{m+1},\ldots,t_{m+n})}
\prod_{j=1}^{m+n}dt_j\;
\omega_1(t_1,\ldots,t_m)
\omega_2(t_{m+1},\ldots,t_{m+n})
\nonumber\\
&&-
\int_{\min(t_{m+1},\ldots,t_{m+n})>\max(t_{1},\ldots,t_{m})}
\prod_{j=1}^{m+n}dt_j\;
\omega_1(t_1,\ldots,t_m)
\omega_2(t_{m+1},\ldots,t_{m+n}).
\label{sp3.3}
\qqq
Obviously, the definition of the iterated commutator~(\ref{sp3.2})
depends on the choice of the zero-point of $t$ parametrization. If
this zero-point is shifted by $\Delta t$, then
\qq
\lim_{\Delta t\rightarrow 0}
\frac{\Delta\int_{\cK}[\omega_1,\omega_2]}{\Delta t}=
2\omega_1(0)\oint_{\cK}\omega_2 - 2\omega_2(0)\oint_{\cK}\omega_1.
\label{f3.2}
\qqq
Therefore if both integrals $\oint_{\cK}\omega_{1,2}$ are equal to
zero, then the integral $\int_{\cK}[\omega_1,\omega_2]$ is well
defined.

The iterated commutator appears in our calculations due to the
following
%%%%%%%%%%%%%%%%%%%%%%%%%%%%%%%%%%
\begin{proposition}
\label{p3.2}
Let $\cK$ be a knot in a manifold $M$, $C_{1,2}$ being the meridian
and parallel on $\partial\Tub(\cK)$. Let $M\p$ be a manifold
constructed by $S$-surgery on $\cK$. Let $\omega$ be a closed form in
$M\setminus\Tub(\cK)$ satisfying a condition
\qq
\oint_{C_1}\omega=1.
\label{sp2.1}
\qqq
Let $\omega_1,\omega_2$ be two closed 1-forms in $M$ and suppose that
\qq
\oint_{C_2}\omega=\oint_{\cK}\omega_1
=\oint_{\cK}\omega_2=0.
\label{sp2.2}
\qqq
Then the forms $\omega_1,\omega_2$ and $\omega$ can be extended into
$M\p$. If the tubular neighborhood $\Tub(\cK)$ is infinitely thin
then
\qq
\int_{\Tub\p(\cK)}\omega_1\wedge\omega_2\wedge\omega=
\frac{1}{2}\int_{\cK}[\omega_1,\omega_2].
\label{sp2.3}
\qqq
\end{proposition}
%%%%%%%%%%%%%%%%%%%%%%%%%%%%%%%%%%%%%
To prove this proposition we introduce the functions $f_1$ and
$f_1\p$ such that
\begin{eqnarray}
df_1&=&\omega_1\;\;\;\;\;\;{\rm inside} \;\;\Tub(\cK),
\label{sp2.4}\\
df_1\p&=& \omega_1\;\;\;\;\;\;{\rm inside} \;\;\Tub\p(\cK)
\label{sp2.5}
\end{eqnarray}
and they coincide on the common boundary:
\qq
\left.f_1\right|_{\partial\Tub(\cK)}=
\left.f\p_1\right|_{\partial\Tub\p(\cK)}.
\label{sp2.6}
\qqq
Then
\qq
\int_{\Tub\p(\cK)}\omega_1\wedge\omega_2\wedge\omega=
\int_{\Tub\p(\cK)}df_1\p\wedge\omega_2\wedge\omega=
\int_{\partial\Tub(\cK)}f_1\omega_2\wedge\omega=
-\oint_{\cK}f_1\omega_2
\label{sp2.7}
\qqq
and
\qq
\int_{\cK}[\omega_1,\omega_2]=
\int_0^1dt_2\,\omega_2(t_2)
\left[\int_{t_2}^1 dt_1\omega_1(t_1)-
\int_0^{t_2}dt_1\omega_1(t_1)\right]=
-2\oint_{\cK}f_1\omega_2.
\label{sp2.8}
\qqq
This proves the proposition.

\noindent
\underline{Milnor's Invariant}
\nopagebreak

Suppose that the following linking numbers are all equal to zero:
\qq
l_{ij}=0,\;\;\;\;1\leq i,j\leq 3
\label{sp.10}
\qqq
(a condition $l_{jj}=0$, $1\leq j\leq 3$ is not necessary to define
the triple Milnor's invariants but it will simplify our formulas).
These conditions allow us to extend the forms
$\omega_1,\omega_2,\omega_3$ into the manifold
$M_{\bar{1}\bar{2}\bar{3}}$. The triple Milnor's linking number of
the link components $\cL_1,\cL_2,\cL_3$ is equal to the intersection
number
\qq
l_{123}^{(\mu)}=\int_{M_{\bar{1}\bar{2}\bar{3}}}
\omega_1\wedge\omega_2\wedge\omega_3.
\label{sp.12}
\qqq
The r.h.s. of this formula can be expressed as an integral in $M$.
Indeed, we may split the integral~(\ref{sp.12}):
\qq
l^{(\mu)}_{123}=\int_{M_{123}}\omega_1\wedge\omega_2\wedge\omega_3
+\sum_{j=1}^3\int_{\Tub\p(\cL_j)}\omega_1\wedge\omega_2\wedge\omega_3.
\label{sp.13}
\qqq
If the tubular neighborhoods $\Tub(\cL_j)$ are infinitely thin then
the first integral in the r.h.s. of eq.~(\ref{sp.13}) is equal to
\qq
\int_{M_{123}}\omega_1\wedge\omega_2\wedge\omega_3=
Y^{(6)}_{123}\equiv\int_{M}\omega_1\wedge\omega_2\wedge\omega_3,
\label{sp.7}
\qqq
while the integrals in the sum can be transformed with the help of
the Proposition~\ref{p3.2}:
\qq
\int_{\Tub\p(\cL_k)}\omega_i\wedge\omega_j\wedge\omega_k=
X^{(7)}_{ij,k}\equiv\frac{1}{2}\int_{\cL_k}[\omega_i,\omega_j],
\;\;\;\;\;\;i\neq j\neq k.
\label{f3.4}
\qqq
Therefore
\qq
l^{(\mu)}_{123}=Y^{(6)}_{123}+X^{(7)}_{12,3}+
X^{(7)}_{31,2}+X^{(7)}_{23,1}
\label{f3.5}
\qqq

\noindent
\underline{Feynman Diagrams}
\nopagebreak

Now we will use the Feynman rules derived in Section~\ref{*2} in
order to calculate the cubic coefficient $l^{(3)}_{123}$.
The Feynman diagram contributions to $l^{(3)}_{123}$ are
depicted in Figs.~6 and~7 up to the permutations.
The diagram of Fig.~6 contains three
propagators~(\ref{f.27}), one cubic vertex~(\ref{f.28}) and tree
vertices~(\ref{f.1029}). Its contribution to the exponent of
eq.~(\ref{5.1}) is equal to
\begin{eqnarray}
\label{f3.6}
\lefteqn{
D^{(6)}_{123}=-2i\pi^2 K \va_1\cdot(\va_2\times\va_3)
\int_M d^3y\;\epsilon^{\nu_1\nu_2\nu_3}
}\\
&&\times
\oint_{\cL_1}dx_1^{\mu_1}\Omega_{\mu_1\nu_1}(x_1,y)
\oint_{\cL_2}dx_2^{\mu_2}\Omega_{\mu_2\nu_2}(x_2,y)
\oint_{\cL_3}dx_3^{\mu_3}\Omega_{\mu_3\nu_3}(x_3,y),
\nonumber
\end{eqnarray}
or, in view of eq.~(\ref{sp1.18}),
\qq
D^{(6)}_{123}=-2i\pi^2 K \va_1\cdot(\va_2\times\va_3)
Y^{(6)}_{123}.
\label{f3.7}
\qqq
A diagram of Fig.~7 consists of two propagators~(\ref{f.27}), two
vertices~(\ref{f.1029}) and one vertex~(\ref{f.2029}). Its
contribution is equal to
\qq
D^{(7)}_{23,1}=-i\pi^2 K \va_1\cdot(\va_2\times\va_3)
\int_{\cL_1}\left[\oint_{\cL_2}dx_2\Omega(x_2,\cdot),
\oint_{\cL_3}dx_3\Omega(x_3,\cdot)\right],
\label{f3.8}
\qqq
or, in view of eq.~(\ref{sp1.18})
\qq
D^{(7)}_{23,1}=-i\pi^2 K \va_1\cdot(\va_2\times\va_3)
X^{(7)}_{23,1}.
\label{f3.9}
\qqq
A corresponding cubic term in the exponent of eq.~(\ref{5.1}) is
\qq
3i\pi K l^{(3)}_{123} \va_1\cdot(\va_2\times\va_3)
\label{f3.10}
\qqq
so that
\qq
l^{(3)}_{123}=-\frac{2}{3}\pi\left(
Y_{123}^{(6)}+X^{(7)}_{12,3}+X^{(7)}_{31,2}+X^{(7)}_{23,1}\right).
\label{f3.11}
\qqq
Comparing this expression with the formula~(\ref{f3.5}) for the
triple linking number we see that
%%%%%%%%%%%%%%%%%%%%%%%%%%%%%%%%%%%%%%%%%%%%
\begin{proposition}
\label{pf3.1}
If the linking numbers $l_{ii},l_{jj},l_{kk},l_{ij},l_{jk},l_{ik}$
are all equal to zero, then the cubic coefficient $\lti$ in the
exponent of Reshetikhin's formula~(\ref{5.1}) is proportional to the
triple Milnor's linking number $l^{(\mu)}_{ijk}$:
\qq
l_{ijk}^{(\mu)}=-\frac{3}{2\pi}\lti.
\label{f3.12}
\qqq
\end{proposition}
%%%%%%%%%%%%%%%%%%%%%%%%%%%%%%%%%%%%
The obvious symmetry
\qq
\lti=-l^{(3)}_{jik}
\label{n3.1}
\qqq
leads to the following relation:
\qq
\sum_{1\leq i,j,k\leq n}
\lti\,
\Tr(\vgs\cdot\va_i)(\vgs\cdot\va_j)(\vgs\cdot\va_k)=
2i
\sum_{1\leq i,j,k\leq n}
\lti\,
(\va_i\times\va_j)\cdot\va_k.
\label{n3.2}
\qqq
After substituting eq.~(\ref{f3.12}) into eq.~(\ref{n.1}) and using
eq.~(\ref{n3.2}) we recover eq.~(\ref{f.37}). This proves the
Conjecture~\ref{cf3.1} for $m=3$.

Note that the contribution~(\ref{f3.9}) of the Feynman diagram of
Fig.~7 is proportional to the integral $X^{(7)}_{23,1}$  of
eq.~(\ref{f3.4}) whose value depends on the choice of the zero-point
it the parametrization $t_1$ of the link component $\cL_1$. However
it follows from eq.~(\ref{f3.2}) that this ambiguity can be
compensated by the change of integration variables~(\ref{sp.1}).

%******************************************************
\nsection{A Quartic Milnor's Linking Number}
\label{*5}
%******************************************************

\noindent
\underline{Preliminaries}
\nopagebreak

We begin with establishing another formula for the triple Milnor's
linking number $l^{(\mu)}_{ijk}$ by splitting the
integral~(\ref{sp.12}) in a different way:
\qq
l^{(\mu)}_{ijk}=
\int_{M_{k,\bar{i}\bar{j}}}\omega_i\wedge\omega_j\wedge\omega_k+
\int_{\Tub\p(\cL_k)}\omega_i\wedge\omega_j\wedge\omega_k.
\label{f5.1}
\qqq
All intersection numbers of the closed 2-form
$\omega_i\wedge\omega_j$ in the manifold $M_{\bar{i}\bar{j}}$ are
equal to zero, so it is exact:
\qq
\omega_i\wedge\omega_j=d\omega_{ij}.
\label{f5.2}
\qqq
Therefore
\qq
\int_{M_{k,\bar{i}\bar{j}}}\omega_i\wedge\omega_j\wedge\omega_k
=\int_{M_{k,\bar{i}\bar{j}}}d\omega_{ij}\wedge\omega_k
=-\int_{\partial\Tub(\cL_k)}\omega_{ij}\wedge\omega_k
=\oint_{\cL_k}\omega_{ij}
\label{f5.3}
\qqq
and
\qq
l^{(\mu)}_{ijk}=\int_{\cL_k}
\left(\omega_{ij}+\frac{1}{2}[\omega_i,\omega_j]\right).
\label{f5.4}
\qqq

If we introduce the functions $f_{i,j},f_{i,j}\p$ satisfying
equations
\begin{eqnarray}
df_{i,j}&=&\omega_i\;\;\;\;\;\;{\rm inside} \;\;\Tub(\cL_j),
\label{sp1.9}\\
df_{i,j}\p&=& \omega_i\;\;\;\;\;\;{\rm inside} \;\;\Tub\p(\cL_j),
\label{sp1.10}\\
\left.f_{i,j}\right|_{\partial\Tub(\cL_j)}&=&
\left.f\p_{i,j}\right|_{\partial\Tub\p(\cL_j)},
\label{sp1.11}
\end{eqnarray}
then the integrals of the iterated commutators can be presented in
the way similar to eq.~(\ref{sp2.8}):
\qq
\int_{\cL_k}[\omega_i,\omega_j]=
-2\oint_{\cL_k}f_{i,k}\omega_j=
2\oint_{\cL_k}f_{j,k}\omega_i=
\oint_{\cL_k}\left(f_{j,k}\omega_i-f_{i,k}\omega_j\right).
\label{f5.5}
\qqq

A 1-form $\omega_{ij}$ is defined by eq.~(\ref{f5.2}) in the manifold
$M_{\bar{i}\bar{j}}$ only up to an addition of a closed form (that
is either $\omega_i$ or $\omega_j$). One possible candidate for
$\omega_{ij}$ can be obtained with the help of the propagator
(1,1)-form $\Omega(x,y)$:
\qq
\omega_{ij}(y)=
\int_{M_{ij}}d^3x\;\Omega(x,y)\wedge\omega_i(x)\wedge\omega_j(x)
-\frac{1}{2}\int_{\cL_i}\left[\Omega(\cdot,y),\omega_j(\cdot)\right]+
\frac{1}{2}\int_{\cL_j}\left[\Omega(\cdot,y),\omega_i(\cdot)\right].
\label{sp1.23}
\qqq
Indeed, it follows from eqs.~(\ref{f.30}),~(\ref{sp3.1})
and~(\ref{f5.5}) that
\begin{eqnarray}
d\omega_{ij}(y)&=&\omega_i(y)\wedge\omega_j(y)-
\int_{\partial M_{ij}}d^2 x\;\tilde{\Omega}(x,y)
\omega_i(x)\wedge\omega_j(x)
\label{f5.6}\\
&&-\frac{1}{2}\int_{\cL_i}[d\tilde{\Omega}(\cdot,y),\omega_j(\cdot)]
-\int_0^1dt^{(i)}\,\delta(y-x(t^{(i)}))
\int_0^{t^{(i)}}dt^{(i)}_j \omega_j(x(t^{(i)}_j))
\nonumber\\
&&+\frac{1}{2}\int_{\cL_j}[d\tilde{\Omega}(\cdot,y),\omega_i(\cdot)]
+\int_0^1dt^{(j)}\,\delta(y-x(t^{(j)}))
\int_0^{t^{(j)}}dt^{(j)}_i \omega_i(x(t^{(j)}_i))
\nonumber\\
&=&\omega_i(y)\wedge\omega_j(y)
-\int_0^1dt^{(i)}\,\delta(y-x(t^{(i)}))
\int_0^{t^{(i)}}dt^{(i)}_j \omega_j(x(t^{(i)}_j))
\nonumber\\
\sshift{
+\int_0^1 dt^{(j)}\,\delta(y-x(t^{(j)}))
\int_0^{t^{(j)}}dt^{(j)}_i \omega_i(x(t^{(j)}_i))
}
\nonumber\\
&&-\int_{\partial\Tub(\cL_i)}d^2 x\,
\tilde{\Omega}(x,y)\omega_i(x)\wedge\omega_j(x)
-\int_{\partial\Tub(\cL_j)}d^2 x\,
\tilde{\Omega}(x,y)\omega_i(x)\wedge\omega_j(x)
\nonumber\\
&&+\oint_{\cL_i}\tilde{\Omega}(\cdot,y)\omega_j(\cdot)
-\oint_{\cL_j}\tilde{\Omega}(\cdot,y)\omega_i(\cdot)
\nonumber\\
&=&
\omega_i(y)\wedge\omega_j(y)
-\int_0^1dt^{(i)}\,\delta(y-x(t^{(i)}))
\int_0^{t^{(i)}}dt^{(i)}_j \omega_j(x(t^{(i)}_j))
\nonumber\\
\sshift{
+\int_0^1 dt^{(j)}\,\delta(y-x(t^{(j)}))
\int_0^{t^{(j)}}dt^{(j)}_i \omega_i(x(t^{(j)}_i)).
}
\nonumber
\end{eqnarray}
The last two terms of this equation are equal to zero when
$y\in M_{ij}$, however they are useful in deriving the formula for
contour integrals of $\omega_{ij}$ by using Stoke's theorem:
%%%%%%%%%%%%%%%%%%%%%%%%%%%%%%%%%%%%%
\begin{proposition}
\label{pf5.1}
Let $C$ be a contour in $M_{ij}$ which is a boundary of a
2-dimensional orientable surface $\Sigma\in M$. Let $\{P_i\}$ and
$\{P_j\}$ be sets of points where $\Sigma$ intersects $\cL_i$ and
$\cL_j$, $\sign{P_i}$  and $\sign{P_j}$ being the signatures of these
intersections. Then for $\omega_{ij}$ given by eq.~(\ref{sp1.23})
\qq
\oint_{C}\omega_{ij}=\int_{\Sigma}\omega_i\wedge\omega_j-
\sum_{P_i}\sign{P_i}\int_0^{t^{(i)}(P_i)}\omega_j+
\sum_{P_j}\sign{P_j}\int_0^{t^{(j)}(P_j)}\omega_i.
\label{f5.7}
\qqq
\end{proposition}
%%%%%%%%%%%%%%%%%%%%%%%%%%%%%%%%%%%%%%%%%%%%%%

Let us choose an infinitely small meridian $C_1^{(i)}(t^{(i)})$
near
the point $t^{(i)}$ as the contour $C$ and a small disk which
intersects $\cL_i$ at the point $t^{(i)}$ as the surface $\Sigma$.
Then the first integral of eq.~(\ref{f5.7}) is infinitely small
because the singularity of the form $\omega_i$ near $\cL_i$ is
compensated by the jacobian measure factor in polar coordinates.
%%%%%%%%%%%%%%%%%%%%%%%%%%%%%%%%%%%%%%%%%%%%%
\begin{corollary}
\label{cf5.1}
If the meridians $C_1^{(i)}(t^{(i)})$ and $C_1^{(j)}(t^{(j)})$ are
infinitely small, then
\qq
\oint_{C_1^{(i)}(t^{(i)})}\omega_{ij}=
-\int_0^{t^{(i)}}\omega_j,\;\;\;\;\;\;
\oint_{C_1^{(j)}(t^{(j)})}\omega_{ij}=
-\int_0^{t^{(j)}}\omega_i.
\label{f5.8}
\qqq
\end{corollary}
%%%%%%%%%%%%%%%%%%%%%%%%%%%%%%%%%%%%%%%%%%%%%%

\noindent
\underline{Milnor's Invariant}
\nopagebreak

Suppose that in addition to the gaussian linking numbers, the triple
Milnor's invariants are also equal to zero:
\qq
l_{ij}=l^{(\mu)}_{ijk}=0,\;\;\;\;\;\;
1\leq i,j,k\leq 4.
\label{sp1.1}
\qqq
Then the 2-forms $\omega_i\wedge\omega_j$ are exact in the manifold
$M_{\bar{1}\bar{2}\bar{3}\bar{4}}$ because the intersection
numbers~(\ref{sp.12}) are equal to zero. Therefore we can introduce
the 1-forms $\omega_{ij}$ of eq.~(\ref{f5.2}) in
$M_{\bar{1}\bar{2}\bar{3}\bar{4}}$. Their values in
$M_{1234}\in M_{\bar{1}\bar{2}\bar{3}\bar{4}}$ may be given by
eq.~(\ref{sp1.23}). At this point we may use
the Massey product (see e.g.
{}~\cite{Ch}, ~\cite{MoRe} and references therein). The following
2-form
\qq
\omega_{ij,k}=\omega_{ij}\wedge\omega_3
-\frac{1}{2}\omega_{ki}\wedge\omega_j-
\frac{1}{2}\omega_{23}\wedge\omega_i
\label{sp1.3}
\qqq
is closed and the integral
\qq
l^{(M)}_{ij,kl}=\int_{M_{\bar{1}\bar{2}\bar{3}\bar{4}}}
\omega_{ij,k}\wedge\omega_l
\label{sp1.4}
\qqq
is a link invariant which is related to the quartic Milnor's linking
number:
\qq
l^{(\mu)}_{ijkl}=\frac{2}{3}(l^{(M)}_{ij,kl}-
l^{(M)}_{jk,il}).
\label{n2.1}
\qqq
An equivalent but more symmetric formula for $l^{(M)}_{ij,kl}$ can
be obtained by integrating by parts:
\begin{eqnarray}
\lefteqn{
l^{(M)}_{ij,kl}=\int_{M_{\bar{1}\bar{2}\bar{3}\bar{4}}}
\omega_{ij,kl},
}
\label{f5.9}\\
\lefteqn{
\omega_{ij,kl}=\frac{1}{2}\left[
\omega_{ij}\wedge\omega_k\wedge\omega_l
+\omega_{kl}\wedge\omega_i\wedge\omega_j
\right.
}
\label{f5.09}\\
&&
\left.-\frac{1}{2}(\omega_{ki}\wedge\omega_j\wedge\omega_l
+\omega_{jl}\wedge\omega_k\wedge\omega_i)
-\frac{1}{2}(\omega_{jk}\wedge\omega_i\wedge\omega_j
+\omega_{il}\wedge\omega_j\wedge\omega_k)\right].
\nonumber
\end{eqnarray}
We will work directly with the invariants $l^{(M)}_{ij,kl}$ because
they have the obvious symmetries:
\qq
l^{(M)}_{ij,kl}=-l^{(M)}_{ji,kl}=
-l^{(M)}_{ij,lk}=l^{(M)}_{kl,ij}.
\label{sp1.5}
\qqq
They also satisfy a Jacobi identity:
\qq
l^{(M)}_{12,34}+l^{(M)}_{31,24}+l^{(M)}_{23,14}=0.
\label{sp1.05}
\qqq
which indicates that the space of quartic invariants of a 4-component
link is 2-dimensional.

Our goal is to express the invariants $l^{(M)}_{ij,kl}$ as integrals
in $M$. We will work with a particular invariant $l^{(M)}_{12,34}$ in
order to simplify our notations. We split the integral~(\ref{f5.9}):
\qq
\int_{M_{\bar{1}\bar{2}\bar{3}\bar{4}}}\omega_{12,34}
=\int_{M_{1234}}\omega_{12,34}+
\sum_{j=1}^4 \int_{\Tub\p(\cL_j)}\omega_{12,34}.
\label{f5.10}
\qqq
The integral $\int_{M_{1234}}\omega_{12,34}$ becomes
$\int_M\omega_{12,34}$ in the limit of infinitely thin tubular
neighborhoods $\Tub(\cL_j)$. For any 3-form
$\omega_{ij}\wedge\omega_k\wedge\omega_l$ we can use
eq.~(\ref{sp1.23}) in order to obtain the formula
\qq
\int_M\omega_{ij}\wedge\omega_k\wedge\omega_l=
Y^{(7)}_{ij,kl}+X^{(10)}_{i,j,kl}-X^{(10)}_{j,i,kl},
\label{f5.11}
\qqq
here
\begin{eqnarray}
Y^{(7)}_{ij,kl}&=&\int_Md^3 y
\left(\int_M d^3
x\,\omega_i(x)\wedge\omega_j(x)\wedge\Omega(x,y)\right)
\wedge\omega_k(y)\wedge\omega_l(y),
\label{f5.12}\\
X^{(10)}_{i,j,kl}&=&\frac{1}{2}\int_M d^3 y
\left(\int_{\cL_j}[\Omega(\cdot,y),\omega_i(\cdot)]\right)
\wedge\omega_k(y)\wedge\omega_l(y).
\label{f5.13}
\end{eqnarray}
As a result,
\qq
\int_{M_{1234}}\omega_{12,34}=
Z^{(0)}_{12,34}-\frac{1}{2}Z^{(0)}_{31,24}-\frac{1}{2}Z^{(0)}_{23,14},
\label{f5.14}
\qqq
here
\qq
Z^{(0)}_{ij,kl}=Y^{(7)}_{ij,kl}+
\frac{1}{2}\left(
X^{(10)}_{i,j,kl}-X^{(10)}_{j,i,kl}+X^{(10)}_{k,l,ij}-
X^{(10)}_{l,k,ij}\right).
\label{f5.15}
\qqq

Next we turn to the integrals over $\Tub\p(\cL_j)$ in
eq.~(\ref{f5.10}). Consider, for example,
$\int_{\Tub\p(\cL_4)}\omega_{12,34}$. After integrating by parts, we
can turn this integral onto a sum
\begin{eqnarray}
\int_{\Tub\p(\cL_4)}\omega_{12,34}&=&I_1 + I_2,
\label{f5.16}\\
I_1&=&\int_{\Tub\p(\cL_4)}
\left(\omega_{12}\wedge\omega_3-
\frac{1}{2}\omega_{31}\wedge\omega_2-
\frac{1}{2}\omega_{23}\wedge\omega_1\right)\wedge\omega_4,
\label{f5.17}\\
I_2&=&\frac{1}{2}\int_{\partial\Tub(\cL_4)}
\left(\omega_{12}\wedge\omega_{34}-
\frac{1}{2}\omega_{31}\wedge\omega_{24}-
\frac{1}{2}\omega_{23}\wedge\omega_{14}\right).
\label{f5.18}
\end{eqnarray}
To calculate $I_1$ we rewrite it as
\begin{eqnarray}
\lefteqn{
I_1=\int_{\Tub\p(\cL_4)}
\left[\left(\omega_{12}-\frac{1}{2}f\p_{1,4}\omega_2
+\frac{1}{2}f\p_{2,4}\omega_1\right)\wedge\omega_3
\right.
}
\label{sp1.12}\\
&&
\left.
-\frac{1}{2}(\omega_{31}+f\p_{1,4}\omega_3)\wedge\omega_2
-\frac{1}{2}(\omega_{23}-f\p_{2,4}\omega_3)\wedge\omega_1\right]\wedge\omega_4.
\nonumber
\end{eqnarray}
All three 1-forms
\qq
\omega_{12}-\frac{1}{2}f\p_{1,4}\omega_2
+\frac{1}{2}f\p_{2,4}\omega_1,\;\;\;
\omega_{31}+f\p_{1,4}\omega_3,\;\;\;
\omega_{23}-f\p_{2,4}\omega_3
\label{sp1.13}
\qqq
are closed. They also satisfy the condition~(\ref{sp2.2}) of the
Proposition~\ref{p3.2} for $\cK=\cL_4$ because of
eqs.~(\ref{f5.4}), ~(\ref{f5.5}) and~(\ref{sp1.1}). Therefore we can
apply the Proposition~\ref{p3.2} to the r.h.s. of eq.~(\ref{sp1.12}):
\qq
I_1=\frac{1}{2}\int_{\cL_4}\left(
\left[\omega_{12}+\frac{1}{2}[\omega_1,\omega_2],\omega_3\right]
-\frac{1}{2}[\omega_{31},\omega_2]
-\frac{1}{2}[\omega_{23},\omega_1]\right).
\label{sp1.15}
\qqq
At the same time, Corollary~\ref{cf5.1} allows us to express $I_2$
also as a contour integral. Since the form $\omega_{ij}$ is
nonsingular in $\Tub(\cL_l)$, then in the limit of infinitely
thin tubular neighborhood
\qq
\int_{\partial\Tub(\cL_l)}\omega_{ij}\wedge\omega_{kl}=
-\int_0^1dt^{(l)}\omega_{ij}(t^{(l)})
\oint_{C_1^{(l)}(t^{(l)})}\omega_{kl}=
-\frac{1}{2}\int_{\cL_l}[\omega_{ij},\omega_k],
\label{f5.19}
\qqq
so that
\qq
I_2=-\frac{1}{4}\int_{\cL_4}\left(
[\omega_{12},\omega_3]-\frac{1}{2}[\omega_{31},\omega_2]-
\frac{1}{2}[\omega_{23},\omega_1]\right).
\label{f5.20}
\qqq
Combining eqs.~(\ref{f5.16}), ~(\ref{sp1.15}) and~(\ref{f5.20}) we
conclude that
\qq
\int_{\Tub\p(\cL_4)}\omega_{12,34}=
\frac{1}{4}\int_{\cL_4}\left(
[\omega_{12},\omega_3]-\frac{1}{2}[\omega_{31},\omega_2]-
\frac{1}{2}[\omega_{23},\omega_1]\right)
+\frac{1}{4}\int_{\cL_4}[[\omega_1,\omega_2],\omega_3],
\label{f5.21}
\qqq
or after substituting eq.~(\ref{sp1.23}) for $\omega_{ij}$,
\qq
\int_{Tub\p(\cL_l)}\omega_{ij,kl}=
Z^{(l)}_{ij,kl}-\frac{1}{2}Z^{(l)}_{ki,jl}
-\frac{1}{2}Z^{(l)}_{jk,il}
+\frac{3}{2}X^{(9)}_{ij,k,l},
\label{f5.22}
\qqq
here
\begin{eqnarray}
Z^{(l)}_{ij,kl}&=&\frac{1}{2}\left(X^{(10)}_{k,l,ij}-
X^{(11)}_{j,i,l,k}+X^{(11)}_{i,j,l,k}\right),
\label{f5.23}\\
X^{(11)}_{i,j,k,l}&=&\frac{1}{4}\int_{\cL_k}
\left[\left(
\int_{\cL_j}[\Omega(\cdot,\ast),\omega_i(\cdot)]
\right),\,\omega_l(\ast)\right],
\label{f5.24}\\
X^{(9)}_{ij,k,l}&=&\frac{1}{6}\int_{\cL_l}[[\omega_i,\omega_j],\omega_k].
\label{f5.25}
\end{eqnarray}
It remains now to put together eqs.~(\ref{f5.9}), ~(\ref{f5.10}),
{}~(\ref{f5.14}) and~(\ref{f5.22}):
\qq
l^{(M)}_{ij,kl}=Z_{ij,kl}-\frac{1}{2}Z_{ki,jl}
-\frac{1}{2}Z_{jk,il}+\frac{3}{2}Z\p_{ij,kl},
\label{f5.26}
\qqq
here
\begin{eqnarray}
Z_{ij,kl}&=&Z^{(0)}_{ij,kl}+Z^{(i)}_{ij,kl}
+Z^{(j)}_{ij,kl}+Z^{(k)}_{ij,kl}+Z^{(l)}_{ij,kl}
\label{f5.27}\\
&=&Y^{(7)}_{ij,kl}+X^{(10)}_{i,j,kl}
-X^{(10)}_{j,i,kl}+X^{(10)}_{k,l,ij}-
X^{(10)}_{l,k,ij}
\nonumber\\
&&-X^{(11)}_{j,i,l,k}+X^{(11)}_{i,j,l,k}-X^{(11)}_{j,i,k,l}
+X^{(11)}_{i,j,k,l},
\nonumber\\
Z\p_{ij,kl}&=&
X^{(9)}_{ij,k,l}-X^{(9)}_{ij,l,k}+X^{(9)}_{kl,i,j}-
X^{(9)}_{kl,i,j}.
\label{f5.28}
\end{eqnarray}

\noindent
\underline{Feynman Diagrams}
\nopagebreak

The Feynman diagrams that contribute to the quartic term
\qq
(\va_i\times\va_j)\cdot(\va_k\times\va_l)
\label{f5.029}
\qqq
are drawn in Figs.~8--11 up to permutations. Their contributions are
easy to calculate with the help of the Feynman rules derived in
Section~\ref{*2}:
\begin{eqnarray}
D^{(8)}_{ij,kl}&=&4i\pi^3 K
(\va_i\times\va_j)\cdot(\va_k\times\va_l)Y^{(7)}_{ij,kl},
\label{f5.29}\\
D^{(9)}_{ij,k,l}&=&4i\pi^3 K
(\va_i\times\va_j)\cdot(\va_k\times\va_l)X^{(9)}_{ij,k,l},
\label{f5.30}\\
D^{(10)}_{i,j,kl}&=&4i\pi^3 K
(\va_i\times\va_j)\cdot(\va_k\times\va_l)X^{(10)}_{i,j,kl},
\label{f5.31}\\
D^{(11)}_{i,j,k,l}&=&-4i\pi^3 K
(\va_i\times\va_j)\cdot(\va_k\times\va_l)X^{(11)}_{i,j,k,l}.
\label{f5.32}
\end{eqnarray}
In eq.~(\ref{f5.30}) we kept only the part of the contribution of the
diagram of Fig.~9 which is proportional to the quartic
term~(\ref{f5.029}). The sum of all these diagrams with appropriate
permutations should be equal to the quartic term in the exponent of
Reshetikhin's formula~(\ref{5.1}):
\qq
4i\pi K\lfi(\va_i\times\va_j)\cdot(\va_k\times\va_l).
\label{f5.33}
\qqq
Therefore comparing eqs.~(\ref{f5.29})--(\ref{f5.32}) with
eqs.~(\ref{f5.27}) and ~(\ref{f5.28}) we conclude that
\qq
\lfi=\pi^2(Z_{ij,kl}+Z\p_{ij,kl}).
\label{f5.34}
\qqq
The coefficients $\lfi$
as defined by this equation, have the symmetry
properties~(\ref{sp1.5}), however they do not satisfy
eq.~(\ref{sp1.05}). This can be fixed with the help of the Jacobi
identity
\qq
(\va_i\times\va_j)\cdot(\va_k\times\va_l)+
(\va_k\times\va_i)\cdot(\va_j\times\va_l)+
(\va_j\times\va_k)\cdot(\va_i\times\va_l)=0,
\label{f5.35}
\qqq
which allows us to add the same quantity to the coefficients $\lti$,
$l^{(4)}_{ki,jl}$ and $l^{(4)}_{jk,il}$ (and the ones obtained by
permutations~(\ref{sp1.5})) without changing the value of the
sum~(\ref{f.38}). Choosing this quantity to be
\qq
-\frac{\pi^3}{3}\left(Z_{ij,kl}+Z_{ki,jl}+Z_{jk,il}\right)
\label{f5.36}
\qqq
we get a new set of coefficients
\qq
\lti=\pi^2\left[
\frac{2}{3}Z_{ij,kl}-\frac{1}{3}Z_{ki,jl}-\frac{1}{3}Z_{jk,il}
+Z\p_{ij,kl}\right],
\label{f5.37}
\qqq
which satisfy the analog of the Jacobi identity~(\ref{sp1.05})
\qq
\lfi+l^{(4)}_{li,jk}+l^{(4)}_{jl,ik}=0
\label{f5.037}
\qqq
in addition to the symmetries
\qq
l^{(4)}_{ij,kl}=-l^{(4)}_{ji,kl}=
-l^{(4)}_{ij,lk}=l^{(4)}_{kl,ij}.
\label{f5.1037}
\qqq
Finally, comparing eqs.~(\ref{f5.37}) and eq.~(\ref{f5.26}) we arrive
at the following
%%%%%%%%%%%%%%%%%%%%%%%%%%%%%%%%%%%%%%
\begin{proposition}
If all the gaussian linking numbers and triple Milnor's invariants
involving the indices $i,j,k,l$ are equal to zero, then the 4th order
coefficients $\lfi$ of the exponent of Reshetikhin's
formula~(\ref{5.1}) normalized by the symmetries~(\ref{f5.1037}) and
Jacobi identities~(\ref{f5.037})  are related to the quartic
Milnor's linking numbers $l^{(\mu)}_{ij,kl}$:
\qq
l^{(\mu)}_{il,kl}=\frac{1}{\pi^2}(\lfi-
l^{(4)}_{jk,il})
\label{f5.38}
\qqq
\end{proposition}
%%%%%%%%%%%%%%%%%%%%%%%%%%%%%%%%%%%%%%%%%%%%%%%%%%%%%%%%

The symmetries~(\ref{f5.1037}) lead to the equation
\qq
\sum_{1\leq j_1,\ldots,j_4\leq n}
l^{(4)}_{j_1j_2j_3j_4}
\Tr(\vgs\cdot\va_{j_1})\cdots(\vgs\cdot\va_{j_4})=
-2
\sum_{1\leq j_1,\ldots,j_4\leq n}
l^{(4)}_{j_1j_2j_3j_4}
(\va_{j_1}\times\va_{j_2})\cdot(\va_{j_3}\times\va_{j_4}).
\label{n2.2}
\qqq
Substituting it together with eq.~(\ref{f5.38}) into the r.h.s. of
eq.~(\ref{n.1}) we obtain eq.~(\ref{f.38}). Thus we verified the
Conjecture~\ref{cf3.1} for $m=4$.

%************************************
\nsection{Discussion}
%************************************
Reshetikhin's formula~(\ref{5.1}) seems to be an excellent tool for
the study of links. It separates the exponent, which is of order
$K^1$, and the preexponential factor, which is at most of order $K^0$,
for future use in stationary phase approximation. It also incorporates
naturally the Feynman diagrams of links and seems to relate them to
Milnor's invariants. This result may not be so surprising in view of
the fact that, as it was demonstrated in~\cite{BN2} and~\cite{Li},
Milnor's linking numbers are Vassiliev's invariants of the link.
The ambiguities of
link Feynman diagrams also seem to match the ambiguities of Milnor's
linking numbers. Both of these ambiguities appear to be
related to the freedom of changing the integration variables in
Reshetikhin's integral (see eqs.~(\ref{sp.1}) and ~(\ref{sp.2})).
It is remarkable that Feynman diagrams of the Chern-Simons theory
are somehow connected to the Massey product.

The set of Feynman rules that we derived in Section~\ref{*2} is not
unique because of Lie algebra Jacobi identities. In fact, it could
possibly be improved in order to incorporate some natural
combinatorial symmetries. This may facilitate a proof of
Conjecture~\ref{cf3.1} about a relation between the exponent of
Reshetikhin's formula and Milnor's linking numbers.

The exponent of Reshetikhin's formula might deserve further study.
Its stationary phase points are in one-to-one correspondence with the
flat connections in the link complement and the determinant of the
second derivatives of the phase is related to their
Reidemeister-Ray-Singer torsion (it is proportional to the
dominant part of the torsion in the limit of small phases of the
holonomies along the meridians of link components).  We use this fact
in~\cite{RoII} in order to calculate the multivariable Alexander
polynomial of the link in the way that generalizes the recent
Melvin-Morton conjecture~\cite{MeMo}.

\section*{Acknowledgements}

I am thankful to N.~Reshetikhin, A.~Vaintrob and O.~Viro for many
useful discussions.

This work was supported by the National Science Foundation
under Grant No. PHY-92 09978.

\end{document}